\documentclass[12pt]{iopart}
\usepackage{iopams}

\expandafter\let\csname equation*\endcsname\relax 
 \expandafter\let\csname endequation*\endcsname\relax 
 \usepackage{amsmath}

\pdfoutput=1

\usepackage{float}
\usepackage[caption = false]{subfig}
\usepackage{graphicx}
\usepackage{epstopdf}

\newcommand{\pts}{$\cal{PT}$}

\begin{document}

\title[Bright solitons in a $\cal{PT}$-symmetric chain]{Bright solitons in a $\cal{PT}$-symmetric chain of dimers}

\author{Omar B. Kirikchi$^{1}$, Alhaji A Bachtiar$^2$ and Hadi Susanto$^1$}

\address{$^1$  Department of Mathematical Sciences, University of Essex, Wivenhoe Park, Colchester, Essex CO4 3SQ, United Kingdom}
\address{$^2$ Department of Mathematics, University of Indonesia, Indonesia}
\ead{hsusanto@essex.ac.uk}

\begin{abstract}
We study the existence and stability of fundamental bright discrete solitons in a parity-time ($\cal{PT}$)-symmetric coupler composed by a chain of dimers, that is modelled by linearly coupled discrete nonlinear Schr\"odinger equations with gain and loss terms. We use a perturbation theory for small coupling between the lattices to perform the analysis, which is then confirmed by numerical calculations. Such analysis is based on the concept of the so-called anti-continuum limit approach. We consider the fundamental onsite and intersite bright solitons. Each solution has symmetric and antisymmetric configurations between the arms. The stability of the solutions is then determined by solving the corresponding eigenvalue problem. We obtain that both symmetric and antisymmetric onsite mode can be stable for small coupling, on the contrary of the reported continuum limit where the antisymmetric solutions are always unstable. The instability is either due to the internal modes crossing the origin or the appearance of a quartet of complex eigenvalues. In general, the gain-loss term can be considered parasitic as it reduces the stability region of the onsite solitons. Additionally, we analyse the dynamic behaviour of the onsite and intersite solitons when unstable, where typically it is either in the form of travelling solitons or soliton blow-ups.
\end{abstract}



\maketitle

\section{Introduction}

A system of equations is \pts-symmetric if it is invariant with respect to combined parity ($\cal{P}$) and time-reversal ($\cal{T}$) transformations. The symmetry is interesting as it forms a particular class of non-Hermitian Hamiltonians in quantum mechanics \cite{mois11}, that may have a real spectrum up to a critical value of the complex potential parameter, above which the system is in the 'broken \pts\ symmetry' phase \cite{bend98,bend99,bend07}.

The most basic configuration having \pts symmetry is a dimer, i.e.\ a system of
two coupled oscillators where one of the oscillators has damping losses and the other one gains energy from external sources. Considerably dimers are also the most important \pts\ systems as the concept of \pts\ symmetry was first realised experimentally on dimers consisting of two coupled optical waveguides \cite{guo09,rute10} (see also the review \cite{such16} for \pts\ symmetry in optical applications). The experiments have been rapidly followed by many other observations of \pts\ symmetry in different branches of physics, from mechanical to electrical analogues (see the review \cite{kono16}).

When nonlinearity is present in a \pts\ system, nontrivial behaviours may emerge that cease to exist in the linear case, such as the presence of blow-up dynamics in the parameter region of the unbroken phase in the linear counterpart 
\cite{pick13,kevr13,bara13}. When nonlinear dimers are put in arrays where elements with gain and loss are linearly coupled to the elements of the same type belonging to adjacent dimers, one can also obtain a distinctive feature in the form of the existence of solutions localized in space as continuous families of their energy parameter \cite{such11}. The system therefore has two arms with each arm described by a discrete nonlinear Schr\"odinger equation with gain or loss. Here, we study the nonlinear localised solutions, which loosely we also refer to as bright discrete solitons, and their stability analytically and numerically.

In the continuous limit, the coupled equations without gain-loss have been studied in \cite{chu93,wrig89,akhm94,malo96}, where it has been shown that the system admits symmetric, antisymmetric and asymmetric solitons between the arms. Unstable asymmetric solutions bifurcate from the symmetric ones through a subcritical symmetry breaking bifurcation, which then become stable after a tangent (saddle-center) bifurcation. When one adds a gain and loss term in each arm, one obtains $\cal{PT}$-symmetric couplers, which have been considered in \cite{drib11,abdu11,drib11b,alex12,bara12,li14}. In the presence of the linear-gain and loss terms, asymmetric solitons cease to exist, while antisymmetric solitons are always unstable \cite{alex12}, even though those with small amplitudes can live long due to weak underlying instability \cite{drib11}. Symmetric solitons can be stable in a similar fashion to those in the system without gain-loss \cite{alex12}.

The stability of bright discrete solitons in \pts-symmetric couplers was discussed in \cite{such11} using variational methods, where it was shown that symmetric onsite solutions can be stable and there is a critical solution amplitude above which the \pts\ symmetry is broken. The case when the polarity of the \pts-symmetric dimers is staggered along the chain is considered in \cite{ambr15}. The same equations without gain and loss were considered in \cite{herr07} where the symmetric soliton loses its stability through the symmetry-breaking bifurcation at a finite value of the energy, similarly to that in the continuous counterpart \cite{chu93,wrig89,akhm94,malo96}. Recently a similar \pts\ chain of dimers with a slightly different nonlinearity was derived \cite{cher16} to describe coupled chains of parametrically driven pendula as a mechanical analogue of \pts-symmetric systems \cite{bend13}. The stability of bright discrete solitons was established through the applications of the Hamiltonian energy and an index theorem. The nonlinear long-time stability of the discrete solitons was also established using the Lyapunov method in the asymptotic limit of a weak coupling between the pendula \cite{cher16b}. 

In this work, we determine the eigenvalues of discrete solitons in \pts-symmetric couplers analytically using asymptotic expansions. The computation is based on the so-called method of weak coupling or anti-continuum limit. The application of the method in the study of discrete solitons was formulated rigorously in \cite{mack94} for conservative systems. It was then applied to \pts-symmetric networks in \cite{kono12,peli14}. However, no explicit expression of the asymptotic series of the eigenvalues for the stability of discrete solitons has been presented before. Here, in addition to the asymptotic limit of weak coupling between the dimers, we also propose to consider expansions in the coefficient of the gain-loss terms. In this case, explicit computations of the asymptotic series of the eigenvalues become possible.

The manuscript is outlined as follows. In Section 2, we present the mathematical model. In Section 3, we use perturbation theory for small coupling to analyse the existence of fundamental localised solutions. Such analysis is based on the concept of the so-called anticontinuum limit approach. 
The stability of the solitons is then considered analytically in Section 4 by solving a corresponding eigenvalue problem. In this section, in addition to small coupling, the expansion is also performed under the assumption of small coefficient of the gain-loss term due to the non-simple expression of the eigenvectors of the linearised operator. The findings obtained from the analytical calculations are then compared with the numerical counterparts in Section 5. We also produce stability regions for all the fundamental solitons numerically. In this section, we present the typical dynamics of solitons in the unstable parameter ranges by direct numerical integrations of the governing equation. 
We present the conclusion in Section 6.


\section{Mathematical model}

The governing equations describing \pts-symmetric chains of dimers are of the form \cite{such11}
\begin{equation}\label{1}
\begin{split}
  &\dot{u}_{n}=i\sigma|u_{n}|^{2}u_{n} + i\epsilon \Delta_{2}u_{n}+\gamma u_{n}+iv_{n},\\ 
  &\dot{v}_{n}=i\sigma|v_{n}|^{2}v_{n} + i\epsilon \Delta_{2}v_{n}-\gamma v_{n}+iu_{n}.
\end{split}  
\end{equation}
The derivative with respect to the evolution variable (i.e., the propagation distance, if we consider their application in fiber optics) is denoted by the overdot, $u_{n}=u_{n}(t)$, $v_{n}=v_{n}(t)$  are complex-valued wave function at site $n\in\mathbb{Z}$, $\epsilon> 0$ is the constant coefficient of the horizontal linear coupling (coupling constant between two adjacent sites), $\Delta_{2}u_{n}=(u_{n+1}-2u_{n}+u_{n-1})$ and $\Delta_{2}v_{n}=(v_{n+1}-2v_{n}+v_{n-1})$ are the discrete Laplacian term in one spatial dimension, the gain and loss acting on complex variables $u_{n}$, $v_{n}$ are represented by the positive coefficient $\gamma$, i.e.\ $\gamma> 0$. The nonlinearity coefficient is denoted by $\sigma$, which can be scaled to $+1$ without loss of generality due to the case of focusing nonlinearity that we consider. 
Bright discrete soliton solutions satisfy the localisation conditions $u_{n},v_{n}\rightarrow 0$ as $n\rightarrow \pm \infty$.

The focusing system has static localised solutions that can be obtained from substituting
\begin{equation}\label{3.1a}
	u_{n}=A_{n}e^{i\omega t},\,\,\
	v_{n}=B_{n}e^{i\omega t}
\end{equation}
into (\ref{1}) to yield the equations
\begin{equation}\label{2}
\begin{split}
\omega A_{n}&=|A_{n}|^{2}A_{n}+\epsilon(A_{n+1}-2A_{n}+A_{n-1})-i\gamma A_{n}+B_{n},\\ 
\omega B_{n}&=|B_{n}|^{2}B_{n}+\epsilon(B_{n+1}-2B_{n}+B_{n-1})+i\gamma B_{n}+A_{n},
\end{split}
\end{equation}
where $A_{n}$, $B_{n}$ are complex-valued and the propagation constant $\omega\in\mathbb{R}$. 

\section{Solutions of weakly coupled equations}
\label{exist}

In the uncoupled limit, i.e.\ when $\epsilon=0$, the chain (\ref{1}) becomes the equations for the dimer. The static equation (\ref{2}) has been analysed in details in \cite{li11,kono12}, where it was shown that there is a relation between $\omega$ and $\gamma$ above which there is no time-independent solution to (\ref{2}) (see also the analysis below). When $\epsilon$ is nonzero, but small enough, the existence of solutions emanating from the uncoupled limit can be shown using the Implicit Function Theorem. The existence analysis of \cite{cher16} can be adopted here despite the slightly different nonlinearity as the Jacobian of our system when uncoupled shares a rather similar invertible structure (see also \cite{kono12,peli14} that have the same nonlinearity in the governing equations but different small coupling terms). However, below we will not state the theorem and instead derive the asymptotic series of the  solutions.

Using perturbation expansion, solutions of the coupler (\ref{2}) for small coupling constant $\epsilon$ can be expressed analytically as
\begin{eqnarray}\label{14}
  A_{n}=A_{n}^{(0)}+\epsilon A_{n}^{(1)}+\dots
  ,\quad B_{n}=B_{n}^{(0)}+\epsilon B_{n}^{(1)}+\dots. 
\end{eqnarray}
By substituting the above expansions into Eq. (\ref{2}) and collecting the terms in successive powers of $\epsilon$, one obtains at $\mathcal{O}(1)$ and $\mathcal{O}(\epsilon)$, respectively, the equations
\begin{eqnarray}\label{15}
 A_{n}^{(0)}=B_{n}^{(0)}(\omega-B_{n}^{(0)}B_{n}^{*{(0)}}-i\gamma),\quad B_{n}^{(0)}=A_{n}^{(0)}(\omega-A_{n}^{(0)}A_{n}^{*{(0)}}+i\gamma),
\end{eqnarray}
and
\begin{equation}
\label{15a}
\begin{split}
 A_{n}^{(1)}&=B_{n}^{(1)}(\omega-2B_{n}^{(0)}B_{n}^{*{(0)}}-i\gamma)-{B_{n}^{(0)}}^2B_{n}^{*(1)}-\Delta_{2}B_{n}^{(0)},\\ 
 B_{n}^{(1)}&=A_{n}^{(1)}(\omega-2A_{n}^{(0)}A_{n}^{*{(0)}}+i\gamma)-{A_{n}^{(0)}}^2A_{n}^{*(1)}-\Delta_{2}A_{n}^{(0)}.
\end{split}
\end{equation}

It is well-known that there are two natural fundamental solutions representing bright discrete solitons that may exist for any $\epsilon$, from the anticontinuum to the continuum limit, i.e.\ an intersite (two-excited-site) and onsite (one-excited-site) bright discrete mode. Here, we will limit our study to these two fundamental modes.

\subsection{Intersite soliton}

In the uncoupled limit, the mode structure $A_{n}^{(0)},\,\,B_{n}^{(0)}$ for the intersite soliton is of the form
\begin{equation}\label{An}
    A_{n}^{(0)}=\left\{
\begin{array}{ccc}
    &\hat{a}_{0}e^{i\phi_{a}}\, &n = 0,1,\\
              &0& \texttt{otherwise},
              \end{array}\right.\quad
    B_{n}^{(0)}=\left\{
\begin{array}{ccc}
    &\hat{b}_{0}e^{i\phi_{b}}\, &n = 0,1,\\
              &0& \texttt{otherwise},
              \end{array}\right.
\end{equation}
with \cite{li11}
\begin{equation}\label{aaa}
 \hat{a}_{0} = \hat{b}_{0}= \sqrt{\omega \mp \sqrt{1-\gamma^2}},\quad \sin(\phi_{b}-\phi_{a})=\gamma,
\end{equation}
which is an exact solution of Eq.\ (\ref{15}). Note that (\ref{aaa}) will have no real solution when $|\gamma|>1$. This is the broken region of \pts-symmetry. The parameter $\phi_{a}$ can be taken as 0, due to the gauge phase invariance of the governing equation (\ref{1}) and henceforth $\phi_{b} = \arcsin\gamma,\,\pi-\arcsin\gamma$. The former phase corresponds to the so-called symmetric configuration between the arms, while the latter is called antisymmetric one. Herein, we also refer to the symmetric and antisymmetric soliton as soliton I and II, respectively. Equation (\ref{aaa}) informs us that $\omega>\sqrt{1-\gamma^2}>0$ and $\omega>-\sqrt{1-\gamma^2}$ are the necessary condition for soliton I and II, respectively.


For the first order correction due to the weak coupling, writing $A_{n}^{(1)}=\tilde{a}_{n,1}e^{i\phi_{a}}$, $B_{n}^{(1)}=\tilde{b}_{n,1}e^{i\phi_{b}}$ and substituting it into equations (\ref{15a}) will yield
\begin{equation}\label{a1b1}
    \tilde{a}_{n,1}=\tilde{b}_{n,1}=\left\{
\begin{array}{ccc}
    &1/(2\hat{a}_{0}) \, &n = 0,1,\\
    &1/\hat{a}_{0}\, &n = -1,2,\\
              &0& \texttt{otherwise}.
              \end{array}\right.
\end{equation}

Equations (\ref{14}),(\ref{An}),(\ref{aaa}),(\ref{a1b1}) are the asymptotic expansion of the intersite solitons. One can continue the same calculation to obtain higher order corrections. Here, we limit ourselves to the first order correction only, which is sufficient to determine the leading order behaviour of the eigenvalues later. 

\subsection{Onsite soliton}

For the onsite soliton, i.e., a one-excited-site discrete mode, one can perform the same computations to obtain the mode structure 
of the form
\begin{equation}\label{An2}
    A_{n}^{(0)}=\left\{
\begin{array}{ccc}
    &\hat{a}_{0}e^{i\phi_{a}}\, &n = 0,\\
              &0& \texttt{otherwise},
              \end{array}\right.\quad
    B_{n}^{(0)}=\left\{
\begin{array}{ccc}
    &\hat{b}_{0}e^{i\phi_{b}}\, &n = 0,\\
              &0& \texttt{otherwise},
              \end{array}\right.
\end{equation}
with (\ref{aaa}). After writing $A_{n}^{(1)}=\tilde{a}_{n,1}e^{i\phi_{a}}$, $B_{n}^{(1)}=\tilde{b}_{n,1}e^{i\phi_{b}}$, the first order correction from (\ref{15a}) is given by
\begin{equation}\label{a1b2}
\tilde{a}_{n,1}=\tilde{b}_{n,1}=\left\{
\begin{array}{ccc}
    &1/\hat{a}_{0}\, &n = 0,\pm1,\\
              &0& \texttt{otherwise}.
              \end{array}\right.
\end{equation}

\section{Stability analysis}

After we find discrete solitons, their linear stability is then determined by solving a corresponding linear eigenvalue problem. To do so, we introduce the linearisation ansatz $u_{n}=(A_{n}+\widetilde{\epsilon}(K_{n}+i L_{n})e^{\lambda t})e^{i\omega t}$, $v_{n}=(B_{n}+\widetilde{\epsilon}(P_{n}+i Q_{n})e^{\lambda t})e^{i\omega t}$, $|\widetilde{\epsilon}| \ll 1$, and substitute this into Eq.\ (\ref{1}) to obtain the linearised equations at $\mathcal{O}(\widetilde{\epsilon}$)
\begin{equation}
\label{8}
\begin{split}
\lambda{K_{n}}&=-(A_{n}^{2}-\omega)L_{n}-\epsilon(L_{n+1}-2L_{n}+L_{n-1})+\gamma K_{n}-Q_{n},\\
\lambda{L_{n}}&=(3A_{n}^{2}-\omega)K_{n}+\epsilon(K_{n+1}-2K_{n}+K_{n-1})+\gamma L_{n}+P_{n},\\
\lambda{P_{n}}&=-(\Re (B_{n})^{2}+3 \Im (B_{n})^{2}-\omega)Q_{n}-\epsilon(Q_{n+1}-2Q_{n}+Q_{n-1})\\
& -(2\Re (B_{n})\Im (B_{n})+\gamma) P_{n}-L_{n},\\
\lambda{Q_{n}}&=(3\Re (B_{n})^{2}+\Im (B_{n})^{2}-\omega)P_{n}+\epsilon(P_{n+1}-2P_{n}+P_{n-1})  \\
& +(2\Re (B_{n})\Im (B_{n})-\gamma)Q_{n}+K_{n}, 
\end{split}
\end{equation}
which have to be solved for the eigenvalue $\lambda$ and the corresponding
eigenvector $[\{K_{n}\},\{L_{n}\},\{P_{n}\},\{Q_{n}\}]^{T}$. As the stability matrix of the eigenvalue problem (\ref{8}) is real valued, $\overline{\lambda}$ and $-{\lambda}$  are also eigenvalues with corresponding eigenvectors $[\{\overline{K_{n}}\},\{\overline{L_{n}}\},\{\overline{P_{n}}\},\{\overline{Q_{n}}\}]^{T}$ and $[\{K_{n}\},\{-L_{n}\},\{P_{n}\},\{-Q_{n}\}]^{T}$ with $\gamma\to-\gamma$, respectively. 
Therefore, we can conclude that the solution $u_{n}$ is (linearly) stable only when $\Re(\lambda)= 0$ for all eigenvalues $\lambda$.

\subsection{Continuous spectrum}

The spectrum of (\ref{8}) will consist of continuous spectrum and discrete spectrum (eigenvalue). To investigate the former, we consider the limit $n\to\pm\infty$, introduce the plane-wave ansatz
$K_{n}=\hat{k}e^{ik_{n}}, L_{n}=\hat{l}e^{ik_{n}}, P_{n}=\hat{p}e^{ik_{n}}, Q_{n}=\hat{q}e^{ik_{n}}
$, $k\in\mathbb{R}$, and substitute the ansatz into (\ref{8}) to obtain
\begin{equation}\label{eig}
\lambda
\left[{\begin{array}{c}  
\hat{k} \\
 \hat{l} \\
   \hat{p}\\
   \hat{q}
   \end{array}}\right] =
\left[{\begin{array}{*{20}c}
   \gamma & \xi &0&-1 \\
   -\xi &\gamma&1&0\\
   0&-1&-\gamma&\xi\\
   1&0&-\xi&-\gamma
\end{array} } \right]
\left[{\begin{array}{c}  
\hat{k} \\
\hat{l} \\
\hat{p} \\
\hat{q}
\end{array}}\right]
\end{equation}
where $\xi=\omega-2\epsilon(\cos k-1)$. The equation 
can be solved analytically to yield the dispersion relation
\begin{eqnarray}
\fl\lambda^2=4\epsilon\omega(\cos k-1)-4\epsilon^2(\cos k-1)^2-\omega^2-1+\gamma^2\pm\left(4\epsilon(\cos k -1)-2\omega\right)\sqrt{1-\gamma^2}.
\label{dps}
\end{eqnarray}
The continuous spectrum is therefore given by $\lambda \in\pm[\lambda_{1{-}}, \lambda_{2{-}}]$ and $\lambda \in\pm[\lambda_{1{+}}, \lambda_{2{+}}]$  with the spectrum boundaries
\begin{eqnarray}
\lambda_{1{\pm}}&=&i\sqrt{1-\gamma^2+\omega^2\pm2\omega\sqrt{1-\gamma^2}},\label{l1}\\
\lambda_{2{\pm}}&=&i\sqrt{1-\gamma^2+8\epsilon\omega+16\epsilon^2+\omega^2+2\sqrt{1-\gamma^2}(\pm\omega-4\epsilon)},\label{l2}
\end{eqnarray}
obtained from (\ref{dps}) by setting $k=0$ and $k=\pi$ in the equation.

\subsection{Discrete spectrum}
\label{sds}

Following the weak-coupling analysis as in Section \ref{exist}, we will as well use similar asymptotic expansions to solve the eigenvalue problem (\ref{8}) analytically, i.e.\ we write
\begin{equation}
\Box = \Box^{(0)}+\sqrt\epsilon \Box^{(1)}+\epsilon\Box^{(2)}+\dots,
\label{box}
\end{equation}
with $\Box=\lambda, K_n, L_n, P_n, Q_n$. We then substitute the expansions into the eigenvalue problem (\ref{8}).

At order $\mathcal{O}(1)$, one will obtain the stability equation for the dimer, which has been discussed for a general value of $\gamma$ in \cite{li11}. The expression of the eigenvalues is simple, but the expression of the corresponding eigenvectors is not, which makes the result of \cite{li11} rather impractical to use. Therefore, here we limit ourselves to the case of small $|\gamma|$ and expand (\ref{box}) further as
\[
\Box^{(j)}=\Box^{(j,0)}+\gamma\Box^{(j,1)}+\gamma^2\Box^{(j,2)}+\dots,
\]
$j=0,1,2,\dots$. Hence, we have two small parameters, i.e.\ $\epsilon$ and $\gamma$, that are independent of each other. For the sake of presentation, the detailed calculations are shown in the Appendix. Below we will only cite the final results.

\subsubsection{Intersite soliton I}
\label{is1}

The intersite soliton I (i.e.\ the symmetric intersite soliton) has three pairs of eigenvalues for small $\epsilon$ and $\gamma$. One pair bifurcate from the zero eigenvalue. They are asymptotically given by
\begin{equation}
\lambda = \sqrt{\epsilon}\left(2\sqrt{\omega-1}+\gamma^2/\left(2\sqrt{\omega-1}\right)+\dots\right)+\mathcal{O}(\epsilon),
\end{equation}
and
\begin{eqnarray}
\fl\lambda=\left\{
\begin{array}{l}
\left(2\sqrt{\omega-2}-\gamma^{2}\frac{\omega-4}{2\sqrt{\omega-2}}+\dots\right)+\epsilon\left(\sqrt{\omega-2}-\gamma^{2}\frac{\omega}{4\sqrt{\omega-2}}
+\dots\right)+\mathcal{O}(\epsilon)^{3/2},\\
\\
\left(2\sqrt{\omega-2}-\gamma^{2}\frac{\omega-4}{2\sqrt{\omega-2}}+\dots\right)+\epsilon\left(\frac{1}{\sqrt{\omega-2}}-\gamma^2\frac{\omega}{4(\omega-2)^{3/2}}+\dots\right)+\mathcal{O}(\epsilon)^{3/2}.
\end{array}
\right.
\end{eqnarray}

\subsubsection{Intersite soliton II}
\label{is2}

The intersite soliton II, i.e.\ the intersite soliton that is antisymmetric between the arms, has three pairs of eigenvalues given by
\begin{equation}
\lambda = \sqrt{\epsilon}\left(2\sqrt{\omega+1}-\gamma^2/\left(2\sqrt{\omega+1}\right)+\dots\right)+\mathcal{O}(\epsilon),
\end{equation}
and
\begin{eqnarray}
\fl\lambda=\left\{
\begin{array}{l}
i\left(2\sqrt{\omega+2}-\gamma^{2}\frac{\omega+4}{2\sqrt{\omega+2}}+\dots\right)\\
-i\epsilon\left(\sqrt{\omega+2}+\gamma^2\frac{3\omega^4+35\omega^3+136\omega^2+208\omega+108}{8\sqrt{\omega+2}\left(\omega^3+6\omega^2+12\omega+8\right)}+\dots\right)+\mathcal{O}(\epsilon)^{3/2},\\
\\
i\left(2\sqrt{\omega+2}-\gamma^{2}\frac{\omega+4}{2\sqrt{\omega+2}}+\dots\right)\\
+i\epsilon\left(\frac{1}{\sqrt{\omega+2}}+\gamma^2\frac{\omega^4+21\omega^3+104\omega^2+184\omega+108}{8\sqrt{\omega+2}\left(\omega^3+6\omega^2+12\omega+8\right)}+\dots\right)+\mathcal{O}(\epsilon)^{3/2}.
\end{array}
\right.
\end{eqnarray}

\subsubsection{Onsite soliton I}
\label{os1}

The onsite soliton has only one eigenvalue for small $\epsilon$ given asymptotically by
\begin{eqnarray}
\fl\lambda=\left(2\sqrt{\omega-2}-\gamma^{2}\frac{\omega-4}{2\sqrt{\omega-2}}+\dots\right)+\epsilon\left(\frac{2}{\sqrt{\omega-2}}-\gamma^{2}\frac{\omega}{2(\omega-2)^{3/2}}+\dots\right)+\dots
\end{eqnarray}

\subsection{Onsite soliton II}
\label{os2}

As for the second type of the onsite soliton, we have
\begin{eqnarray}
\fl\lambda=i\left(2\sqrt{\omega+2}-\gamma^{2}\frac{\omega+4}{\sqrt{\omega+2}}+\dots\right)+2i\epsilon\left(\frac{1}{\sqrt{\omega+2}}-\gamma^{2}\frac{\omega}{(\omega+2)^{3/2}}+\dots\right)+\dots
\end{eqnarray}

\section{Numerical results}

We have solved the steady-state equation (\ref{2}) numerically using a Newton-Raphson method and analysed the stability of the numerical solution by solving the eigenvalue problem (\ref{8}). Below we will compare the analytical calculations obtained above with the numerical results.

\begin{figure}[tbhp!]
\centering
$\begin{array}{ccc}
\includegraphics[width=6cm]{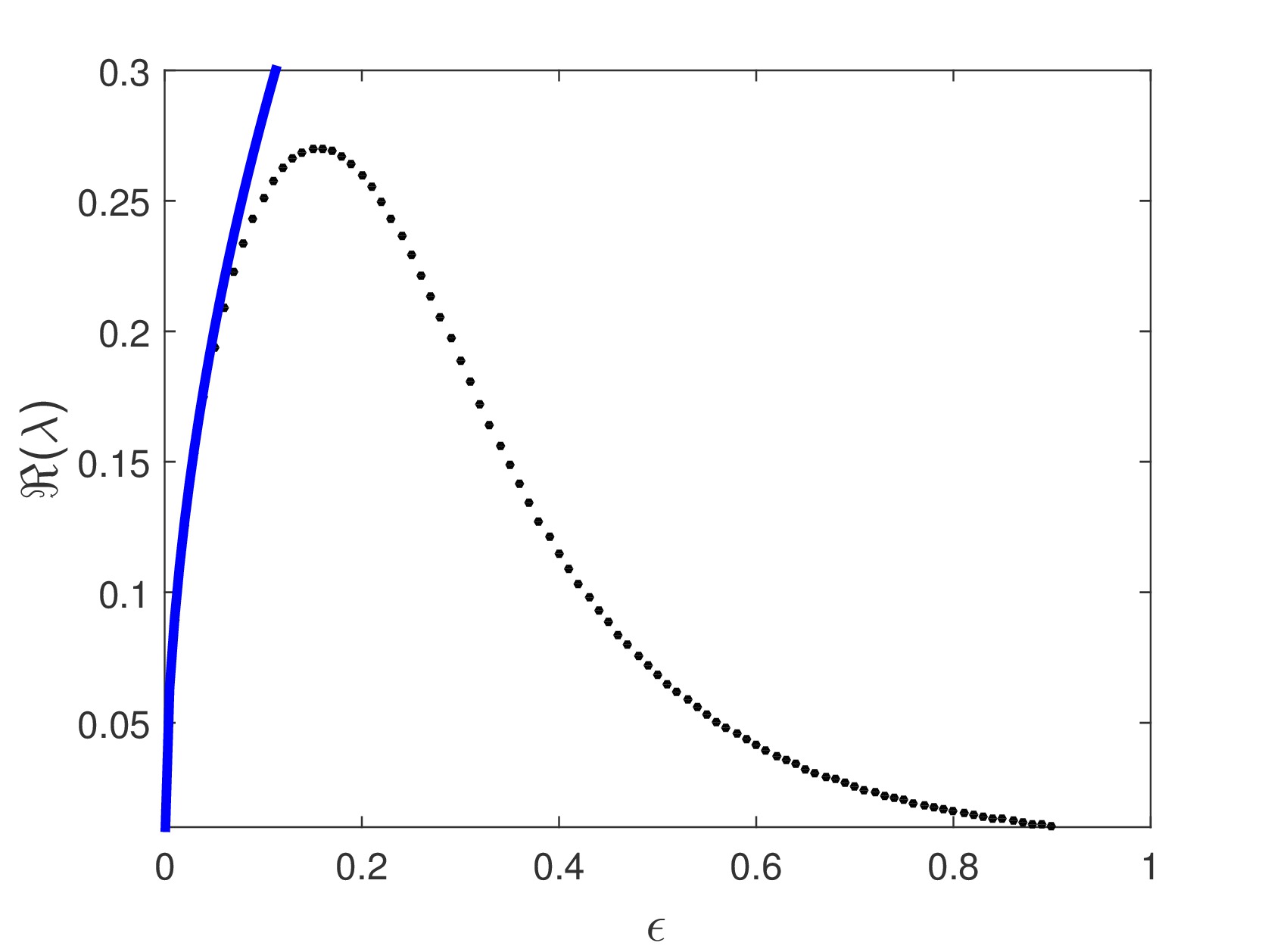}&
\includegraphics[width=6cm]{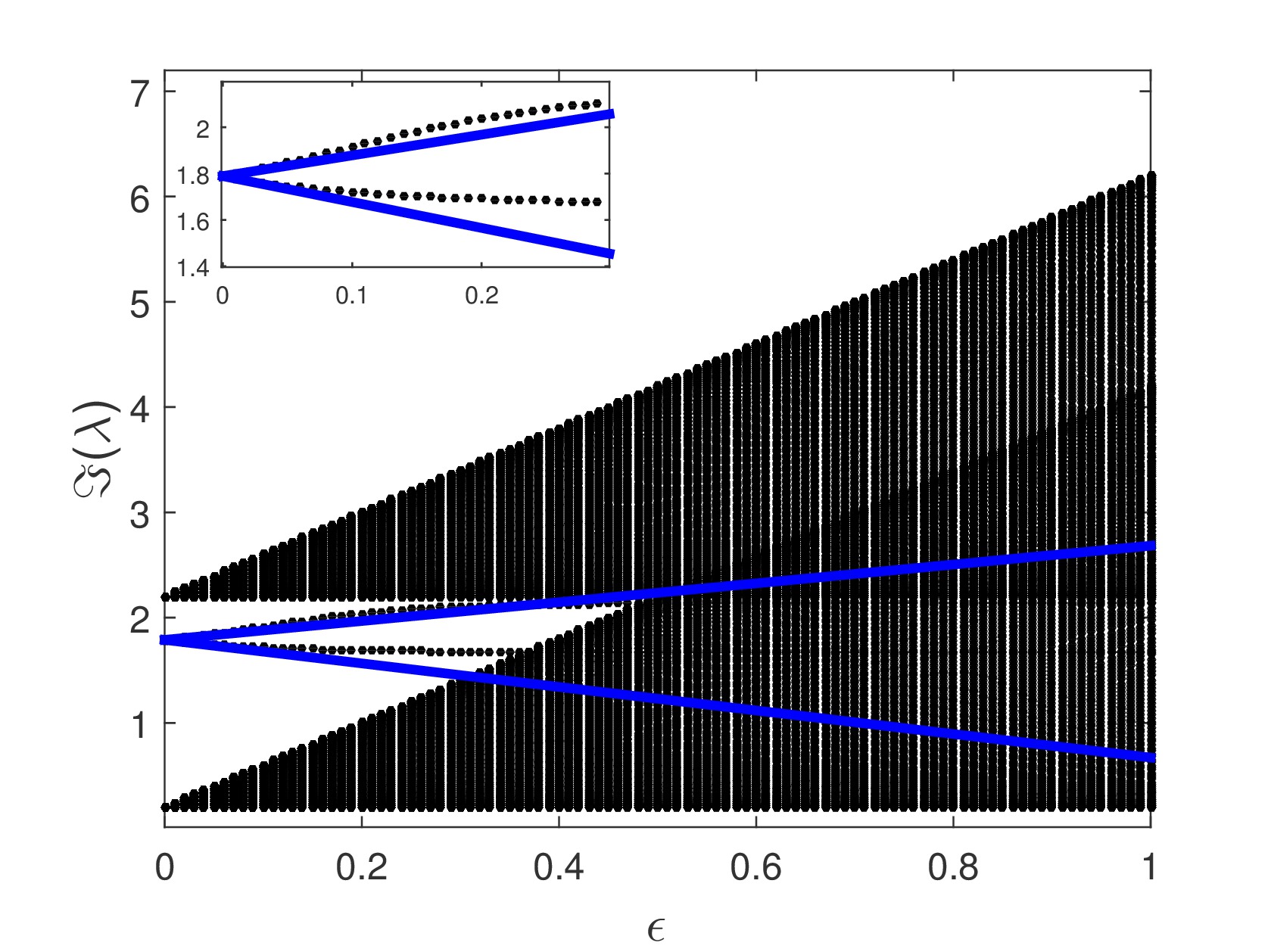}\\
\includegraphics[width=6cm]{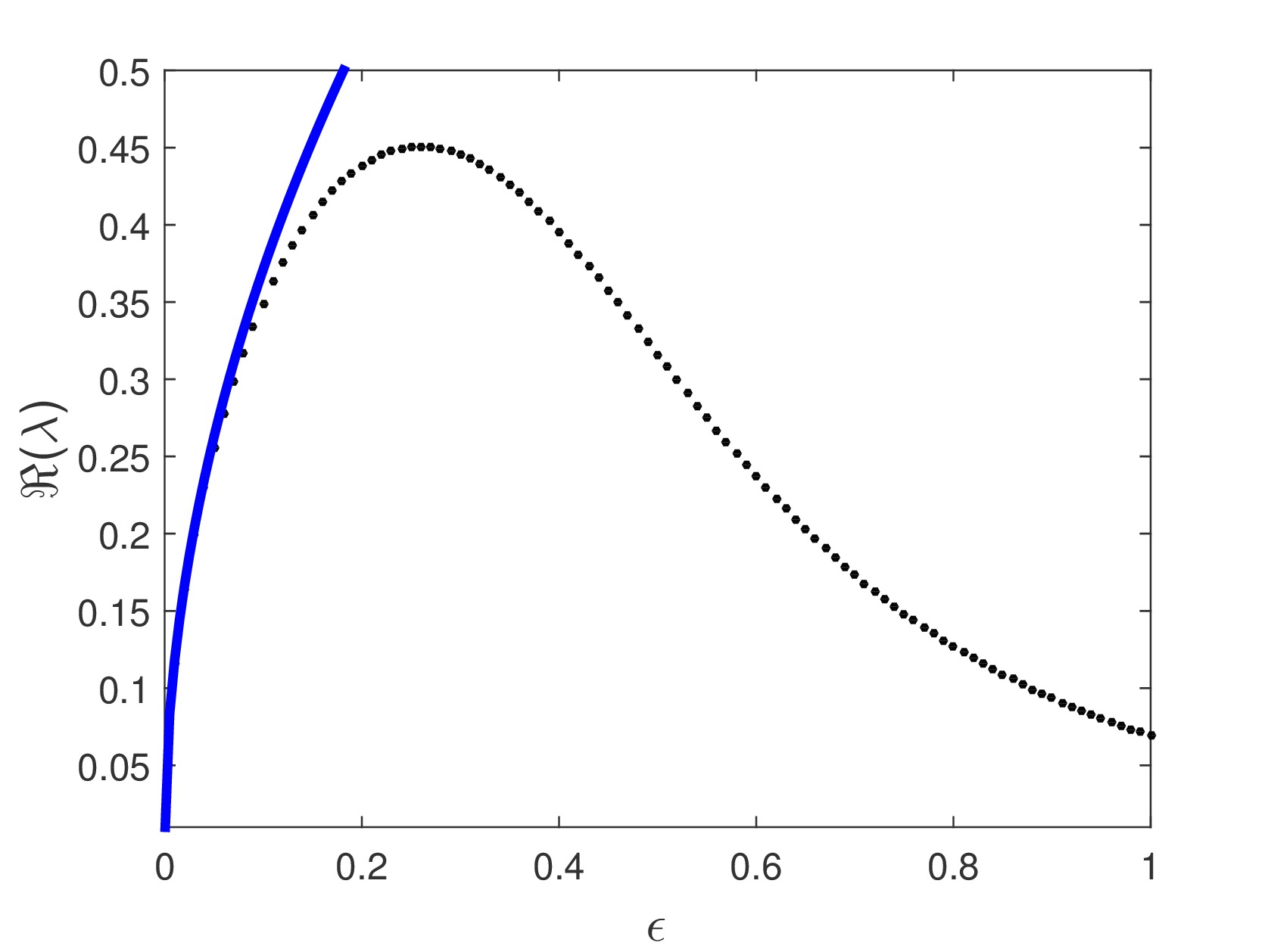}&
\includegraphics[width=6cm]{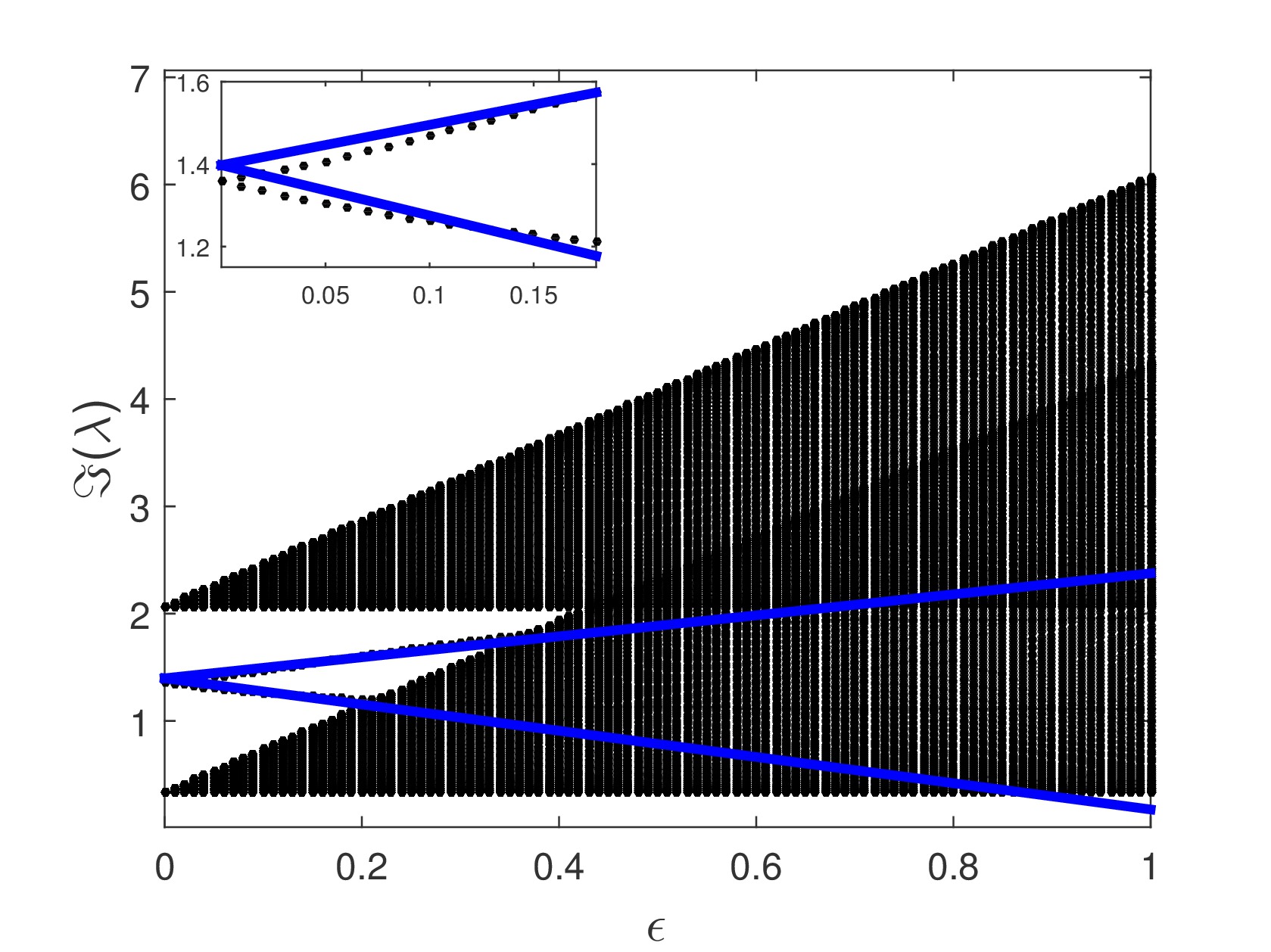}
\end{array}$
\caption{Eigenvalues of intersite soliton I with $\omega=1.2,\,\gamma=0$ (top panels) and $0.5$ (bottom panels). Dots are from the numerics and solid lines are the asymptotic approximations in Section \ref{is1}. The collection of dots forming black regions in the right column corresponds to the continuous spectrum.} \label{fig.converge1}
\end{figure}

\begin{figure}[tbhp!]
\centering
$\begin{array}{ccc}
\includegraphics[width=6cm]{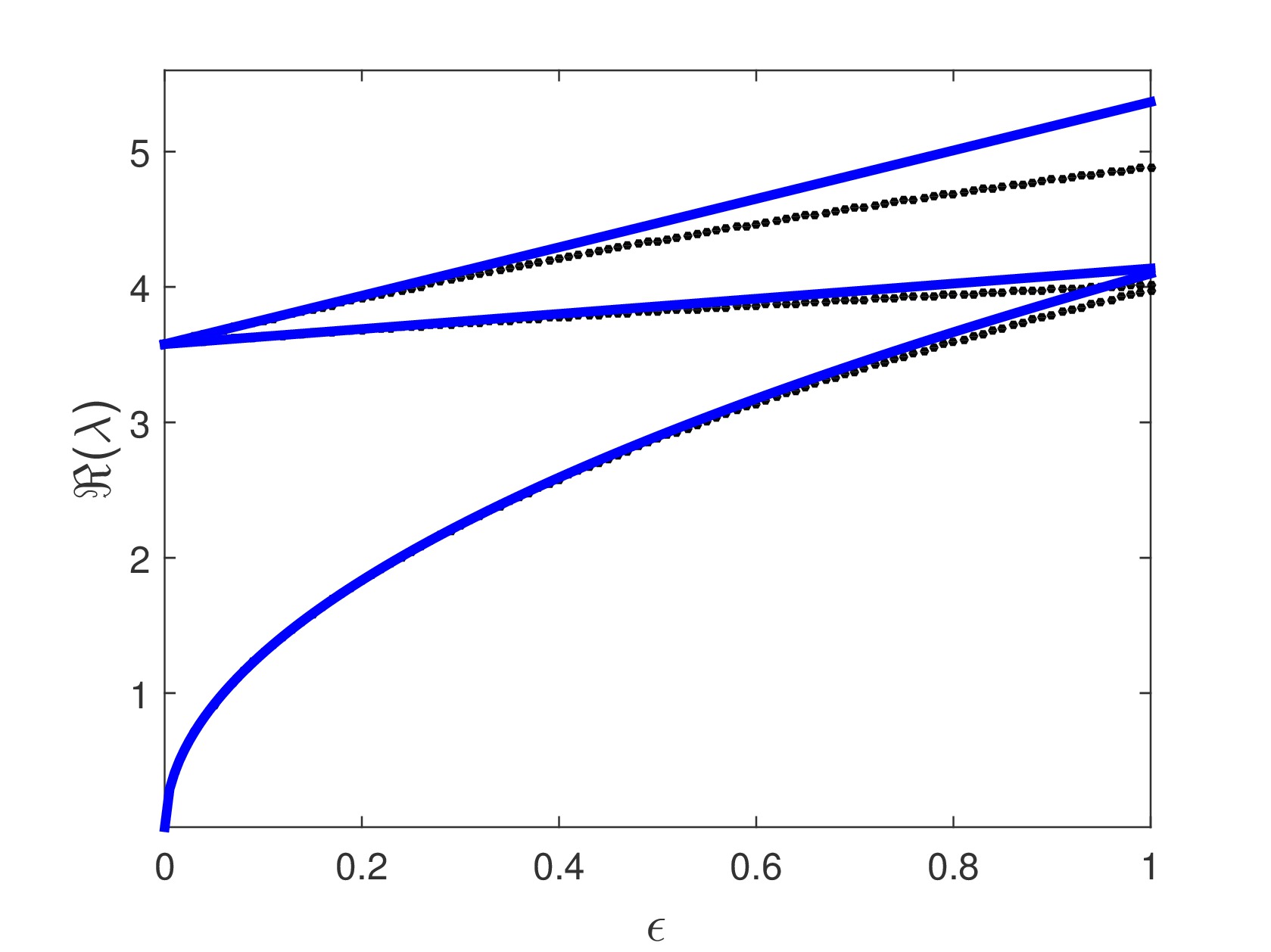}&
\includegraphics[width=6cm]{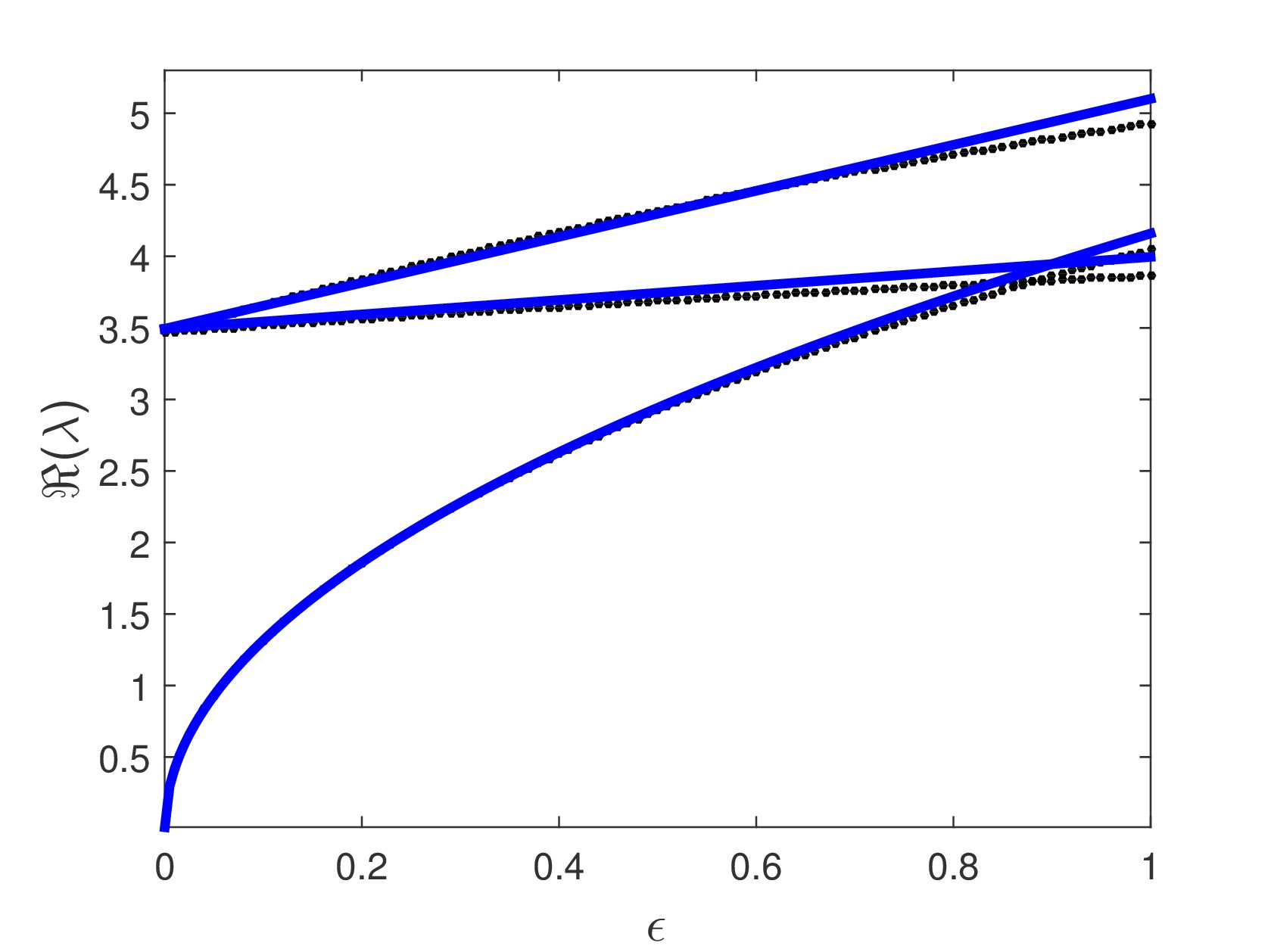}
\end{array}$
\caption{The same as in Fig.\ \ref{fig.converge1} with $\gamma=0$ (left) and $\gamma=0.5$ (right), but for $\omega=5.2$. In this case, all the eigenvalues are real.} \label{fig.converge2}
\end{figure}

First, we consider the discrete intersite soliton I. We show in Fig.\ \ref{fig.converge1} the spectrum of the soliton as a function of the coupling constant $\epsilon$ for $\omega=1.2$ and $\gamma=0,0.5$. On the real axis, one can observe that there is only one unstable eigenvalue that bifurcates from the origin. As the coupling increases, the bifurcating eigenvalue enters the origin again when $\epsilon\to\infty$. Hence, in that limit we obtain a stable soliton I (i.e.\ a stable symmetric soliton). The dynamics of the non-zero eigenvalues as a function of the coupling constant is shown in the right panels of the figure, where one can see that the eigenvalues are on the imaginary axis and simply enter the continuous spectrum as $\epsilon$ increases.

In Fig.\ \ref{fig.converge2}, we plot the eigenvalues for $\omega$ large enough. Here, in the uncoupled limit, all the three pairs of eigenvalues are on the real axis. As the coupling increases, two pairs go back toward the origin, while one pair remains on the real axis (not shown here). In the continuum limit $\epsilon\to\infty$, we therefore obtain an unstable soliton I (i.e.\ an unstable symmetric soliton).

In both figures, 
we also plot the approximate eigenvalues in solid (blue) curves, where good agreement is obtained for small $\epsilon$.

From numerical computations, we conjecture that if in the limit $\epsilon\to0$ all the nonzero eigenvalues $\lambda$ satisfy $\lambda^2>\lambda_{1-}^2$ (see (\ref{l1}-\ref{l2})), then we will obtain unstable soliton I in the continuum limit $\epsilon\to\infty$. However, when in the anticontinuum limit $\epsilon\to0$ all the nonzero eigenvalues $\lambda$ satisfy $\lambda_{1+}^2<\lambda^2<\lambda_{2-}^2$, we may either obtain a stable or an unstable soliton I in the continuum limit.

\begin{figure}[htbp!]
\centering
$\begin{array}{ccc}
\includegraphics[width=4.5cm]{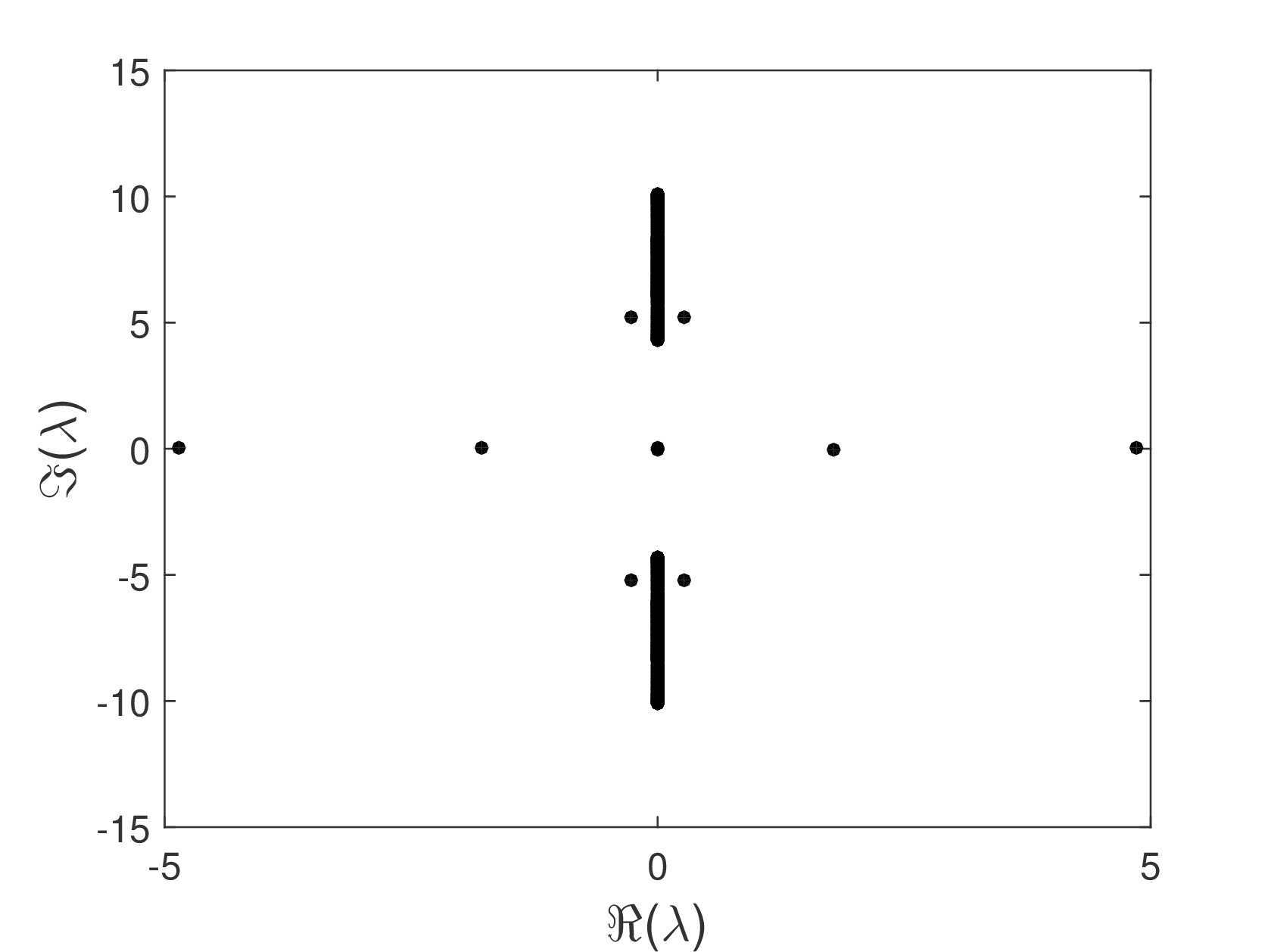}&
\includegraphics[width=4.5cm]{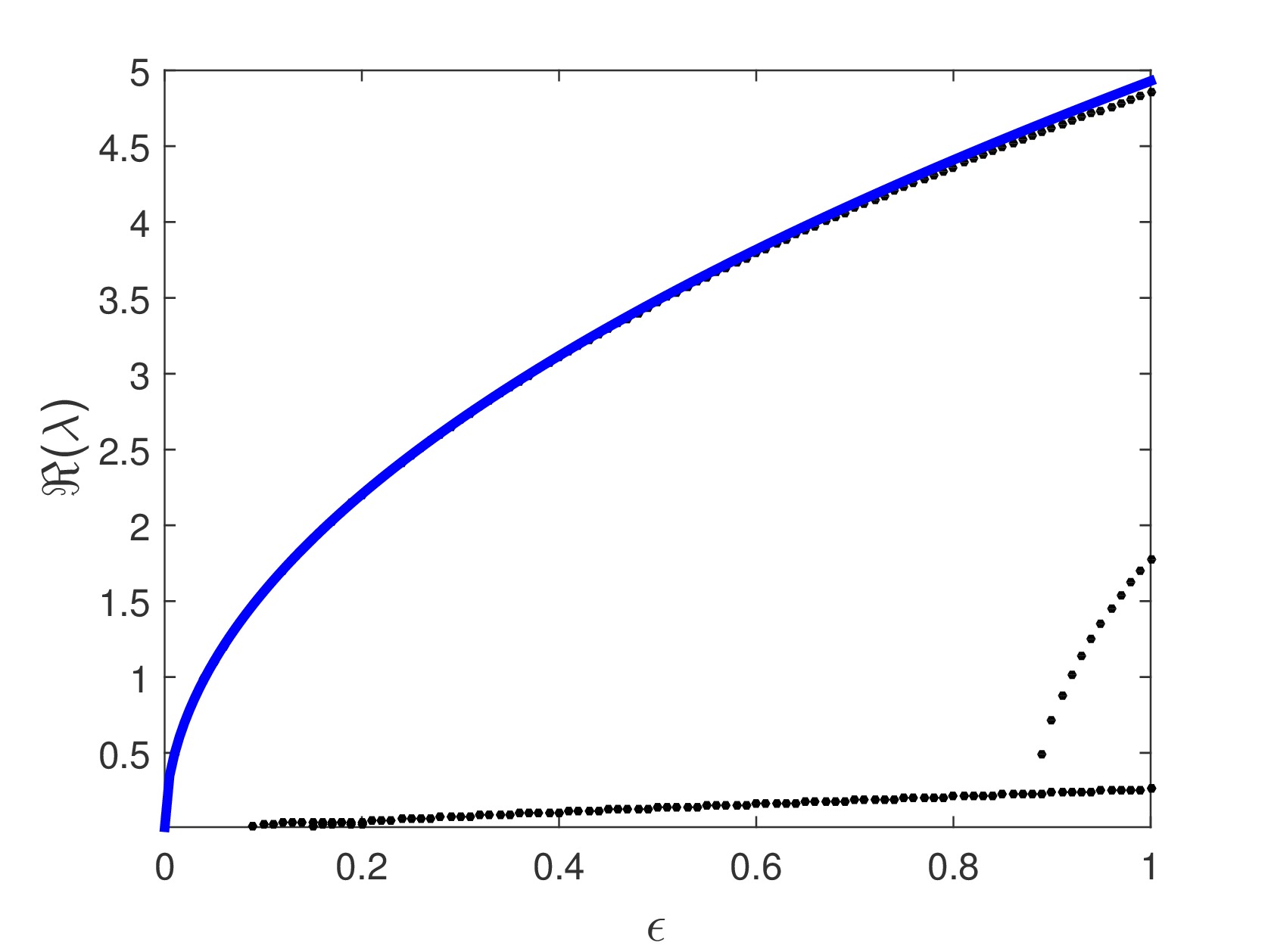}&
\includegraphics[width=4.5cm]{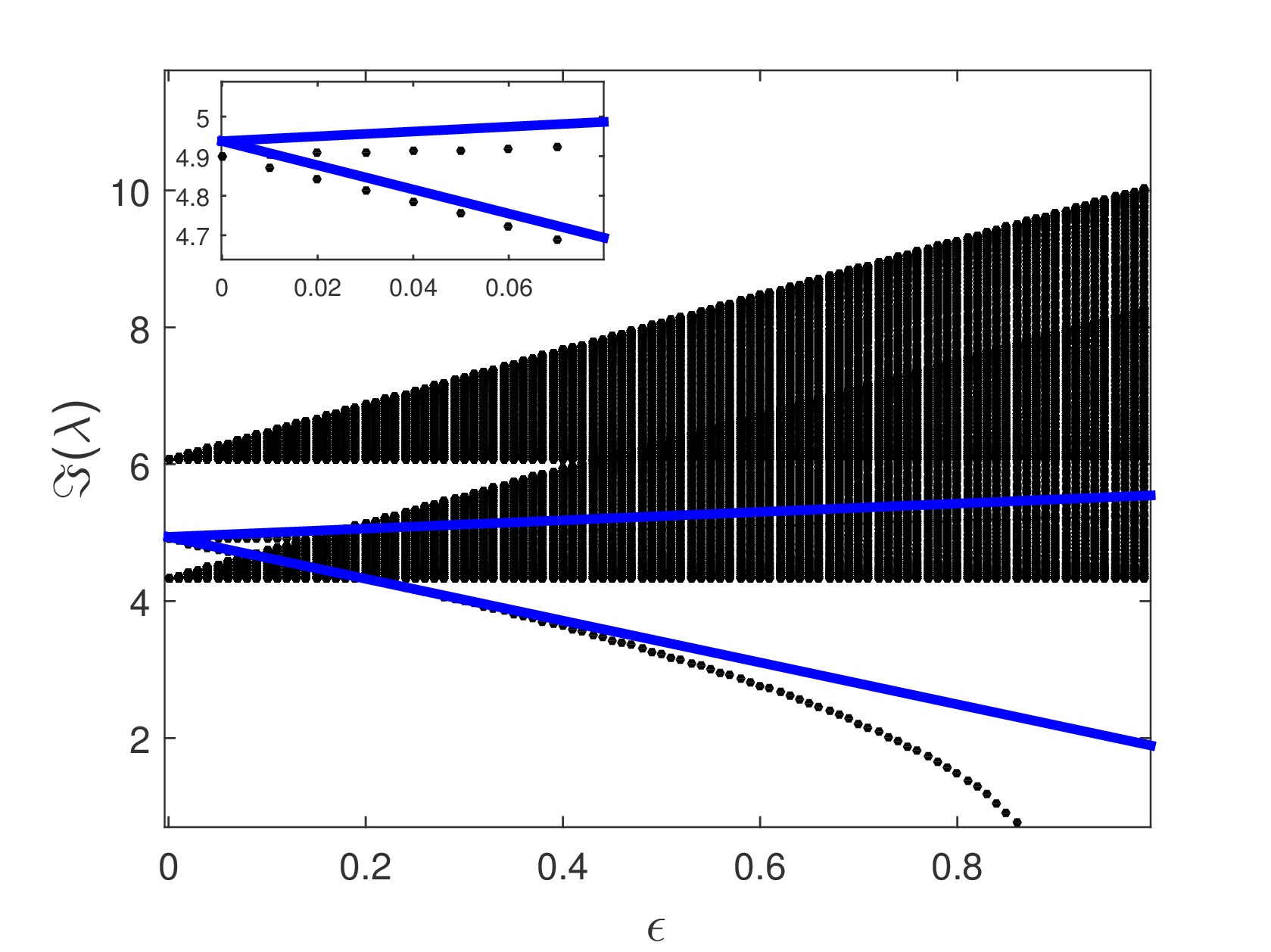}\\
\includegraphics[width=4.5cm]{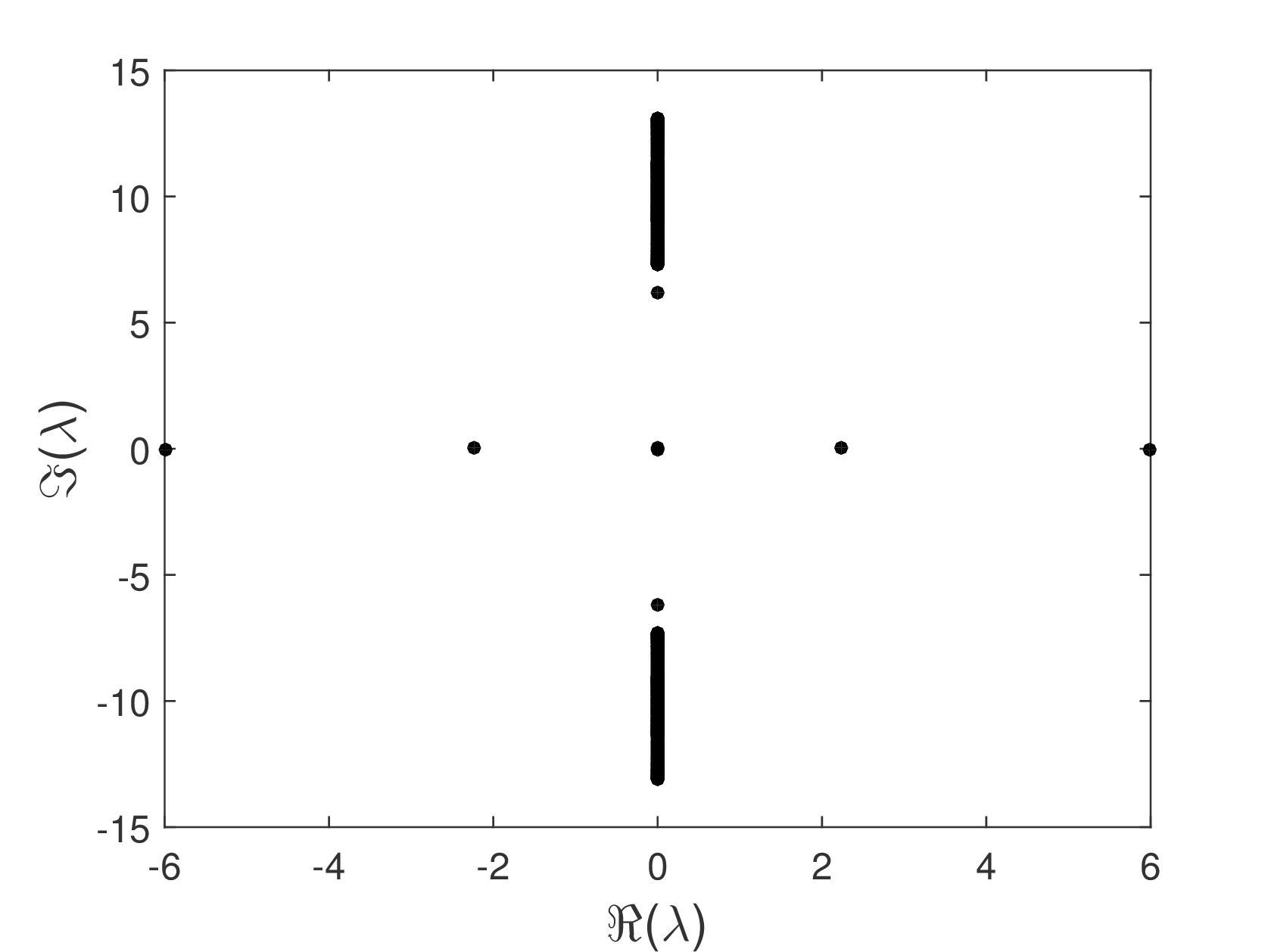}&
\includegraphics[width=4.5cm]{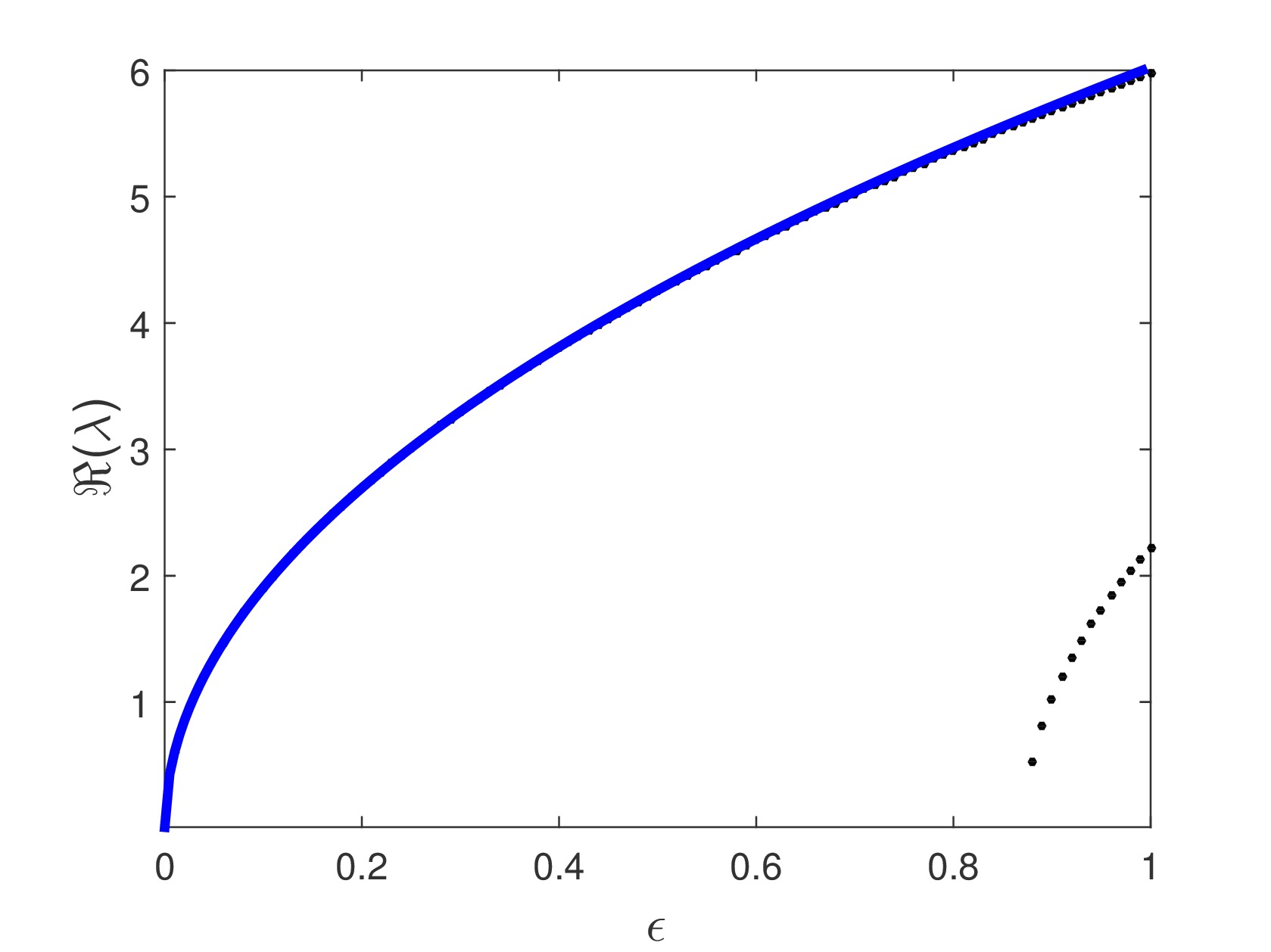}&
\includegraphics[width=4.5cm]{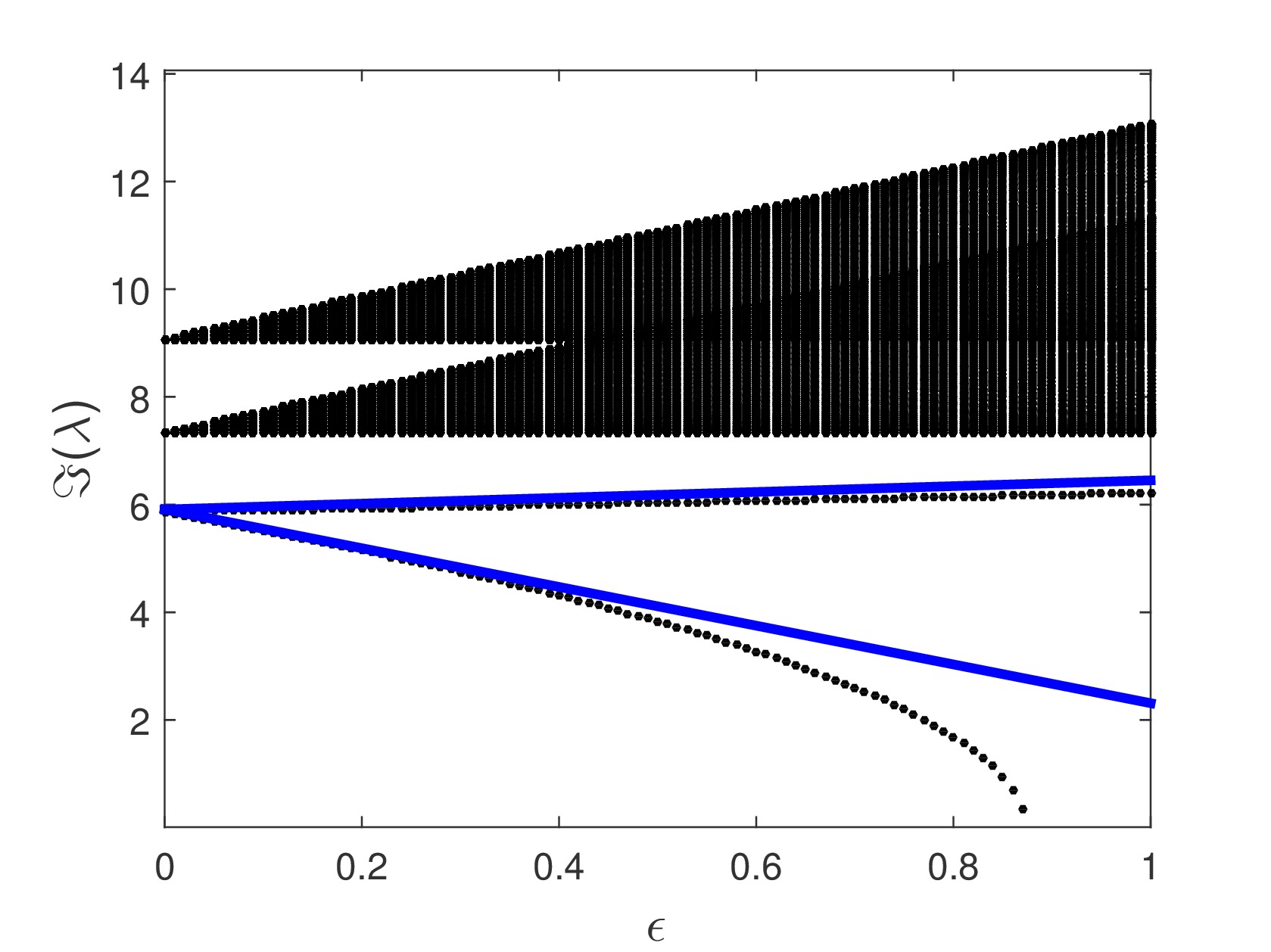}
\end{array}$
\caption{The spectra of intersite soliton II with $\omega=5.2$ (top) and $8.2$ (bottom) and $\gamma=0.5$. The most left panels are the spectra in the complex plane for $\epsilon=1$. Middle and right panels present the eigenvalues as a function the coupling constant. Solid blue curves are the asymptotic approximations.} \label{fig.converge4}
\end{figure}

Next, we consider intersite solitons II (i.e.\ antisymmetric intersite solitons). Shown in Fig.\ \ref{fig.converge4} is the spectrum of the discrete solitons for two values of $\omega$. In both cases, there is an eigenvalue bifurcating from the origin. For the smaller value of $\omega$ (the top panels of the figure), we have the condition that all the nonzero eigenvalues $\lambda$ satisfy $\lambda^2<\lambda_{2-}^2$ in the anticontinuum limit $\epsilon\to0$. The collision between the eigenvalues and the continuous spectrum as the coupling increases creates complex eigenvalues. In the second case using larger $\omega$ (lower panels of the figure), the nonzero eigenvalues $\lambda$ satisfy $\lambda^2>\lambda_{1-}^2$ when $\epsilon=0$. Even though not seen in the figure, the collision between one of the nonzero eigenvalues and the continuous spectrum also creates a pair of complex eigenvalues. Additionally, in the continuum limit both values of $\omega$ as well as the other values of the parameter that we computed for this type of discrete solitons yield unstable solutions.

\begin{figure}[htbp!]
\centering
$\begin{array}{cc}
\includegraphics[width=6cm]{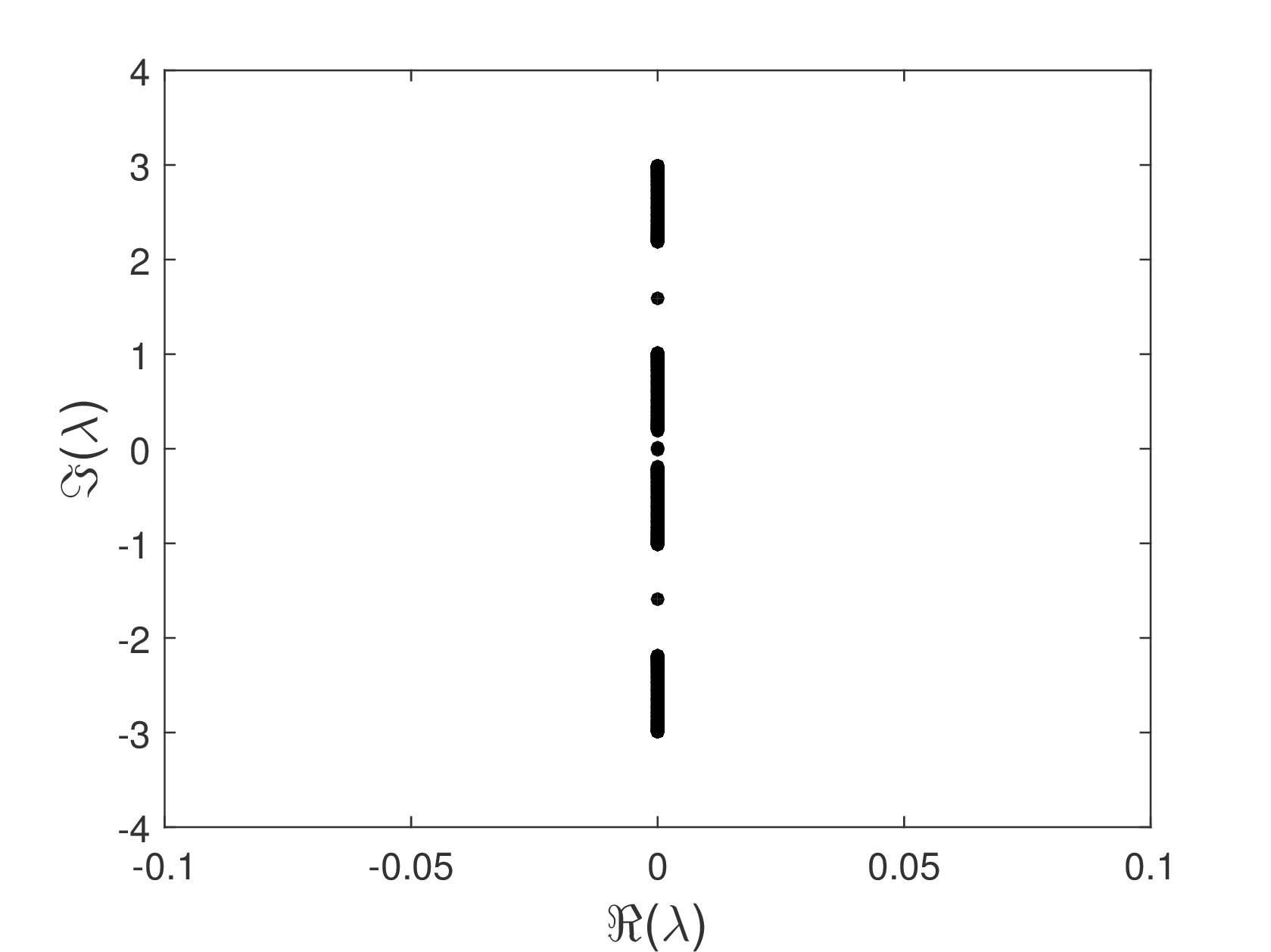}&
\includegraphics[width=6cm]{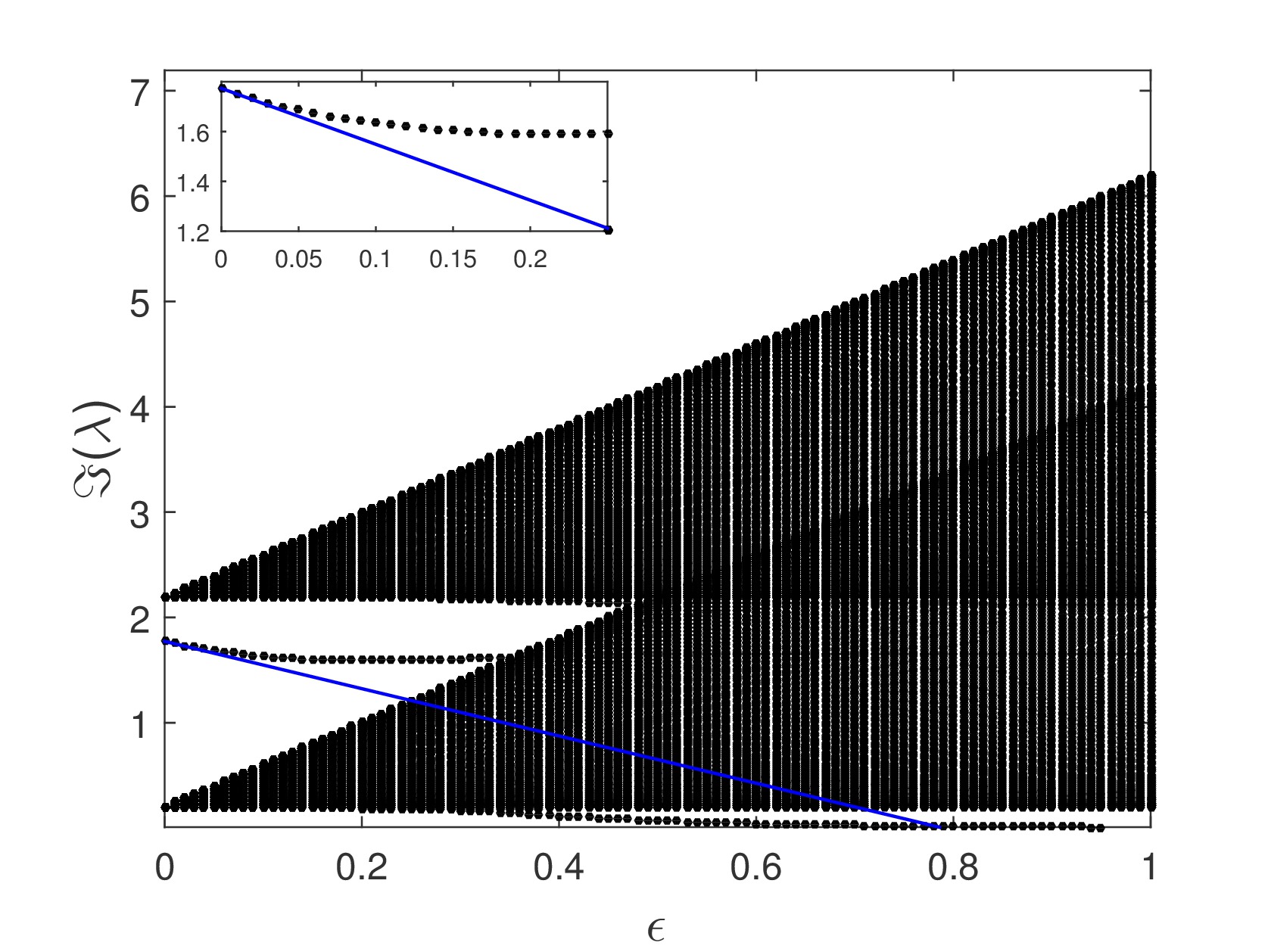}\\
\includegraphics[width=6cm]{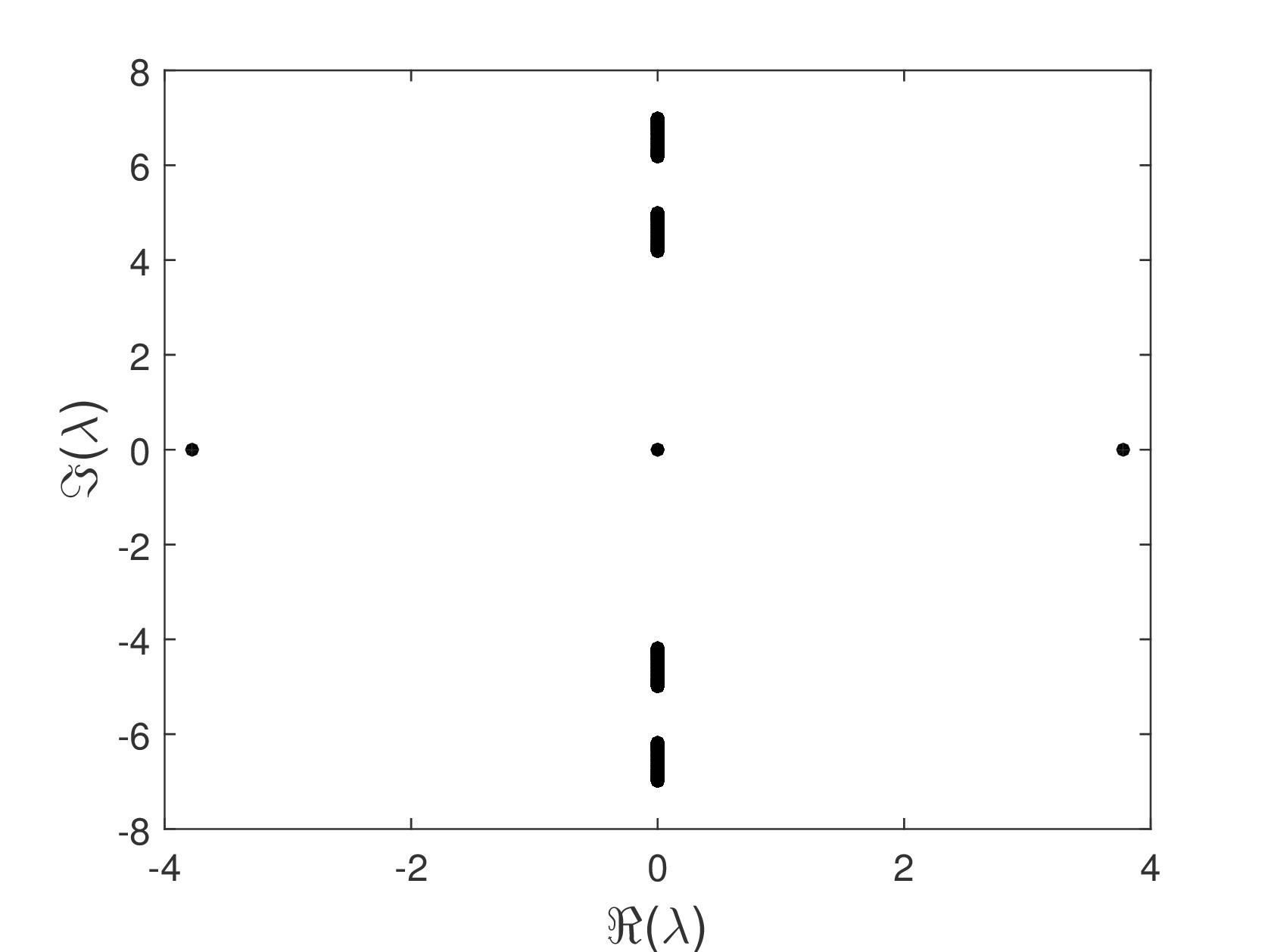}&
\includegraphics[width=6cm]{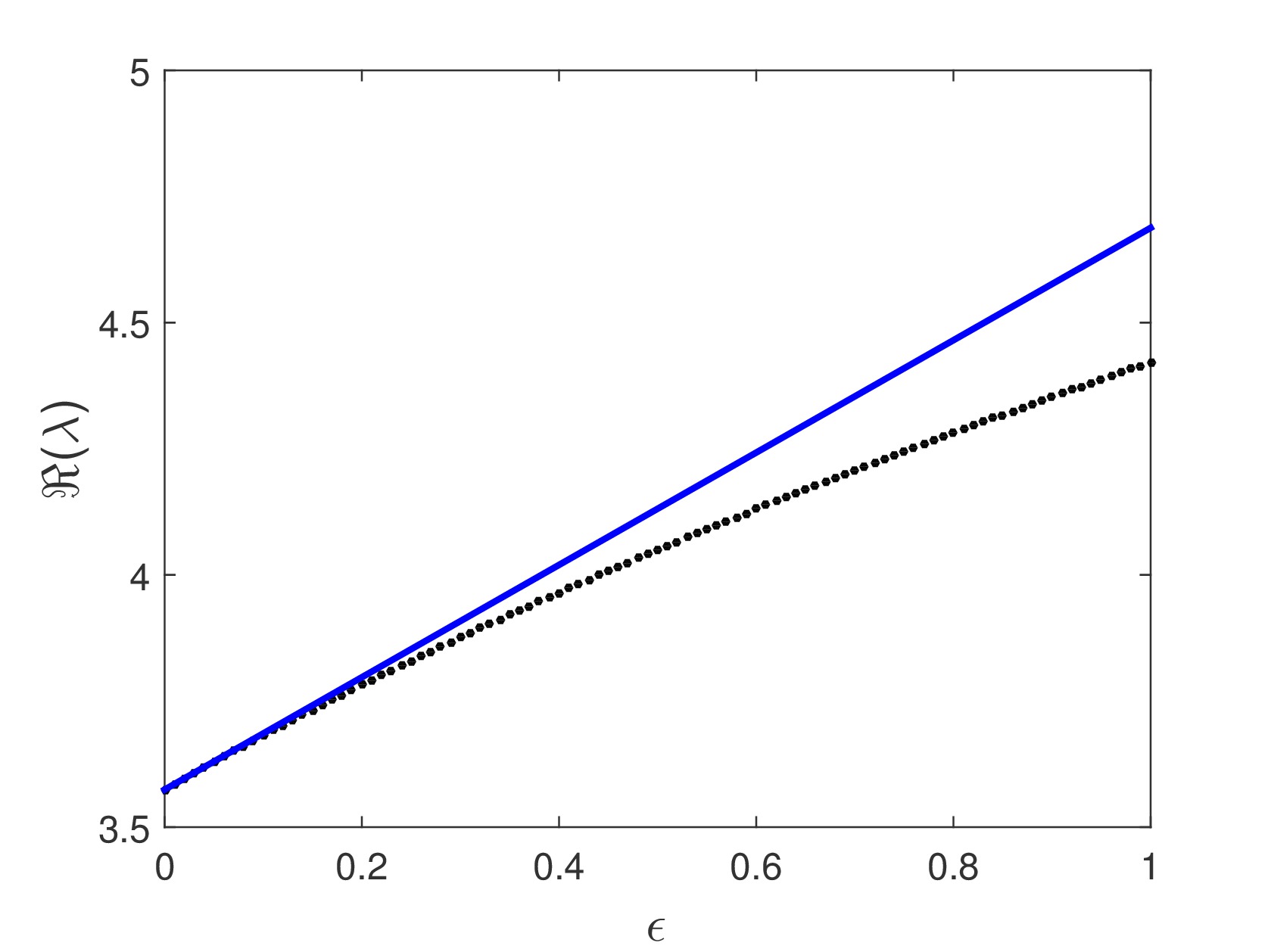}
\end{array}$
\caption{Left panels depict the spectrum of onsite soliton I in the complex plane for $\epsilon=0.2$. Right panels show the eigenvalue as a function of the coupling and its approximation from Section \ref{os1}. Top and bottom panels are for $\omega=1.2$ and $5.2$, respectively. Here, $\gamma=0.1$.} \label{fig.converge5}
\end{figure}

\begin{figure}[htbp!]
\centering
$\begin{array}{cc}
\includegraphics[width=6cm]{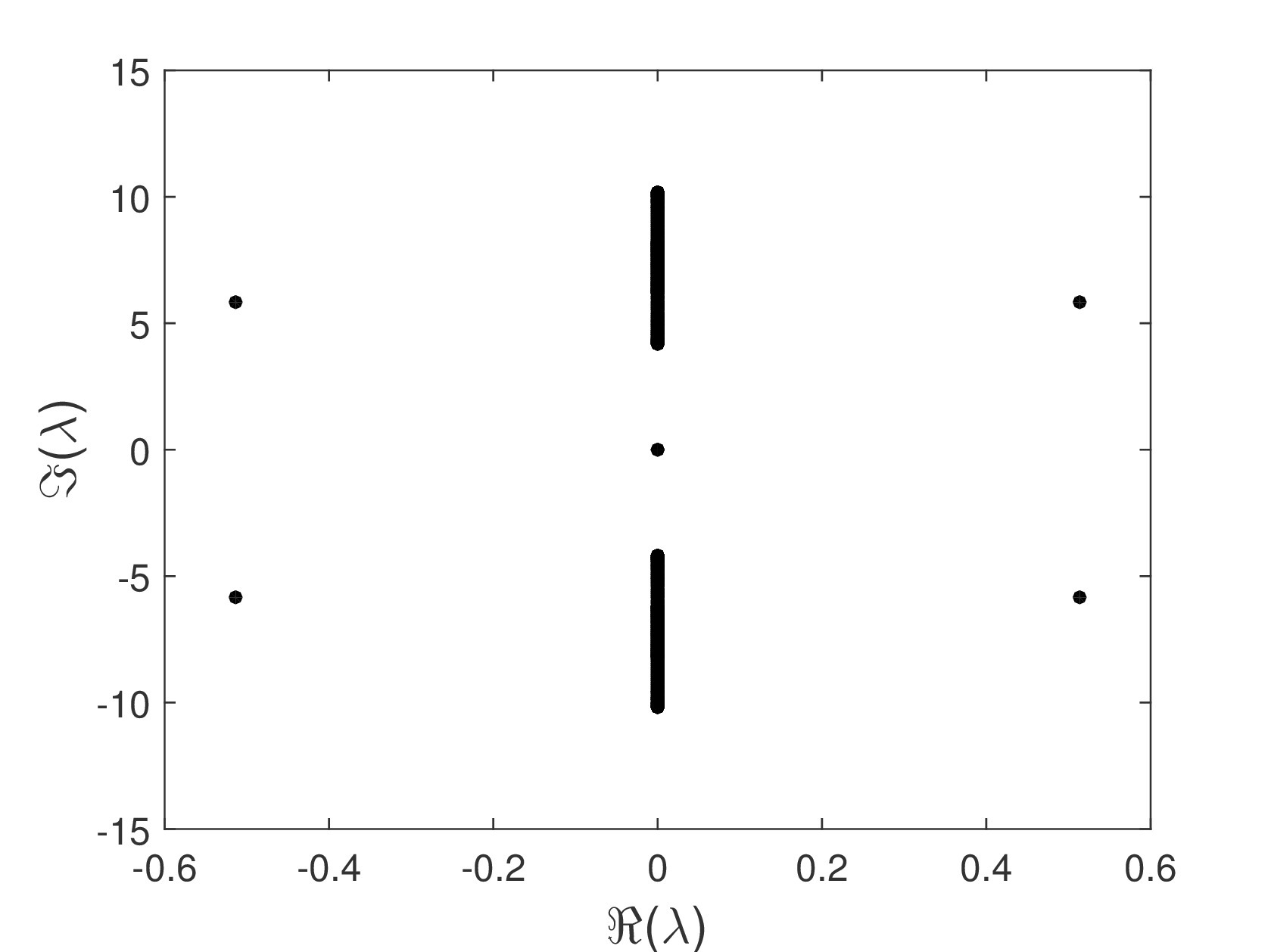}&
\includegraphics[width=6cm]{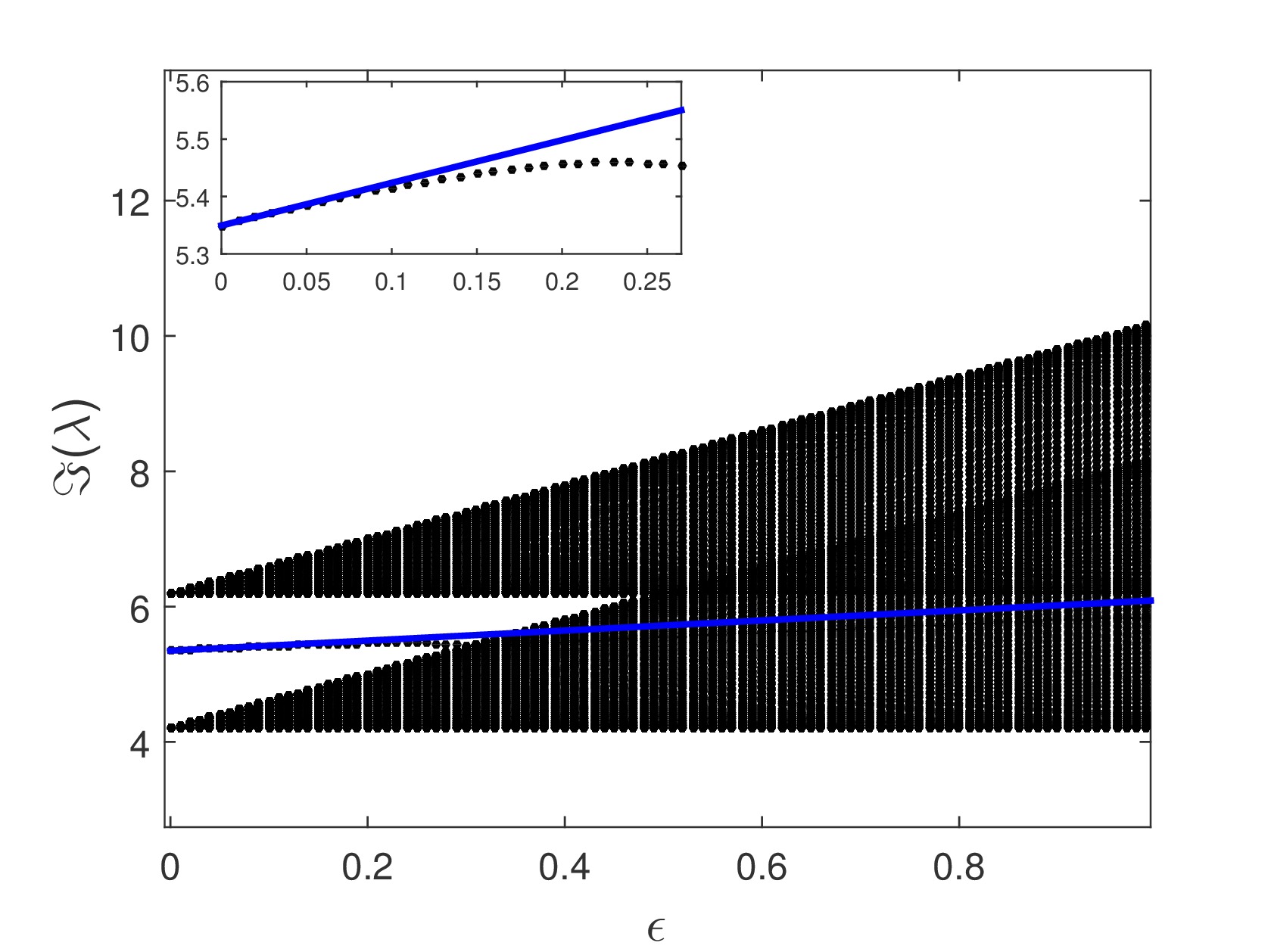}\\
\includegraphics[width=6cm]{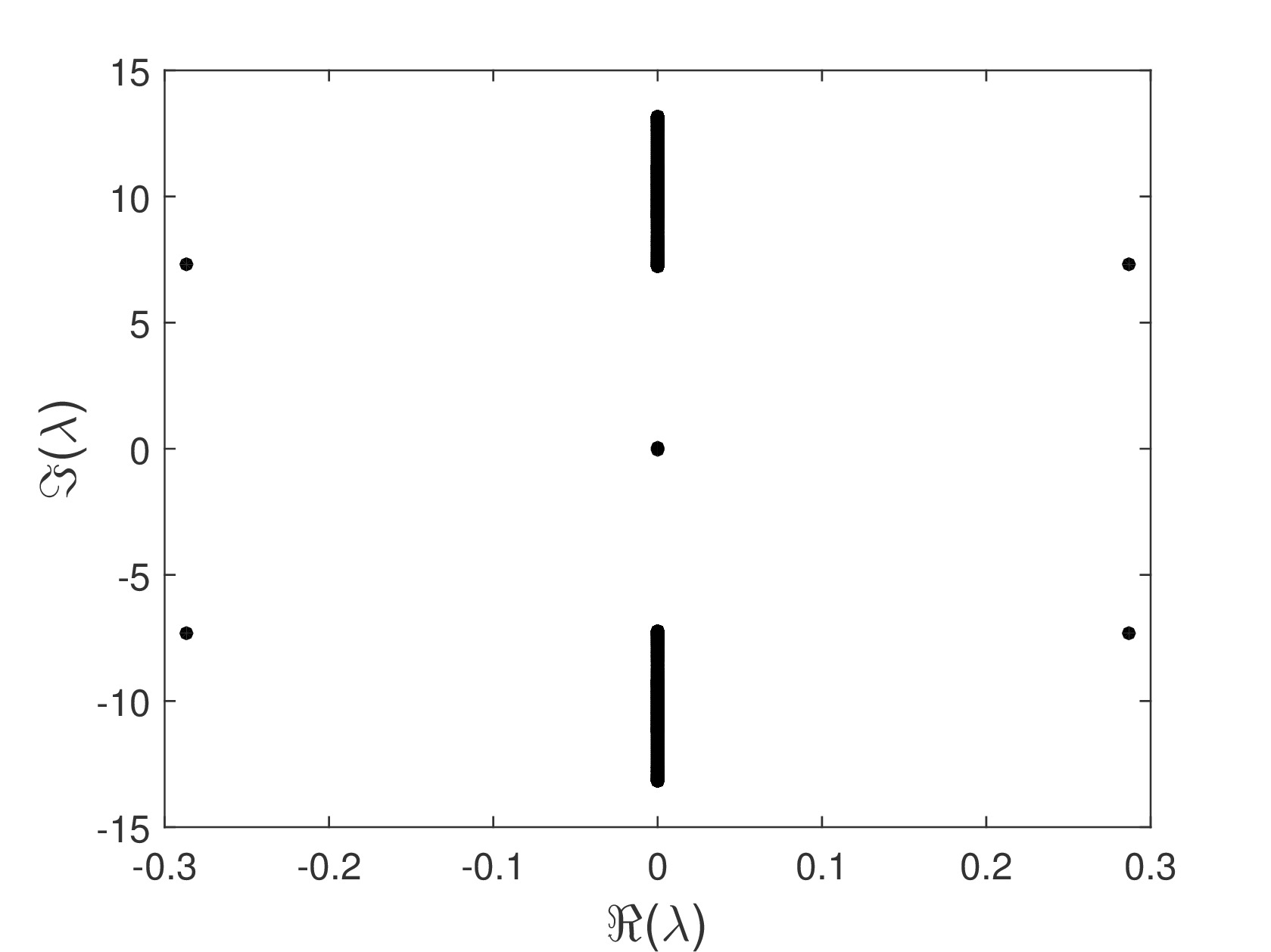}&
\includegraphics[width=6cm]{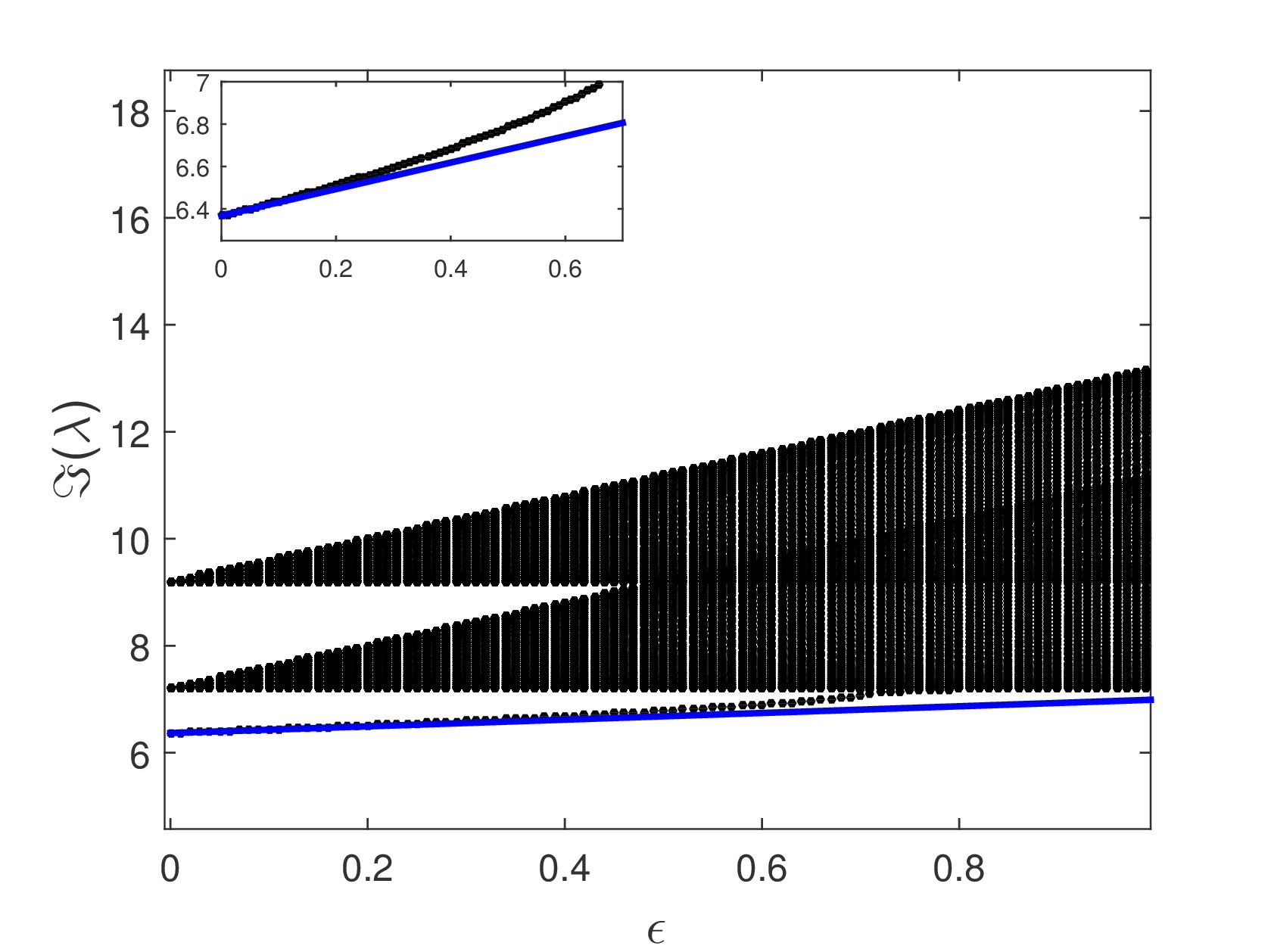}
\end{array}$
\caption{The same as in Fig.\ \ref{fig.converge5}, but for onsite soliton II with $\omega=5.2$ (top) and $8.2$ (bottom). The left panels are with $\epsilon=1$.} \label{fig.converge6}
\end{figure}

We also study onsite solitons. Shown in Figs.\ \ref{fig.converge5} and \ref{fig.converge6} is the stability of discrete solitons type I and II, respectively.

In Fig.\ \ref{fig.converge5}, the top left panel shows that for $(\omega-\sqrt{1-\gamma^2})$ small enough we will obtain stable discrete solitons. For coupling constant $\epsilon$ small, we indeed show it through our analysis depicted as the blue solid line. Numerically we obtain that this soliton is also stable in the continuum limit $\epsilon\to\infty$. However, when $\omega$ is large enough compared to $\sqrt{1-\gamma^2}$, even though initially in the uncoupled limit the nonzero eigenvalue $\lambda$ satisfies $\lambda^2<\lambda_{2-}^2$, one may obtain an exponential instability (i.e.\ instability due to a real eigenvalue). The bottom left panel shows the case when the discrete soliton is already unstable even in the uncoupled limit due to the nonzero eigenvalue that is already real-valued.

Fig.\ \ref{fig.converge6} shows that the antisymmetric solitons are generally unstable due to a quartet of complex eigenvalues, as shown in the left panels of the figure. When the coupling is increased further, there will be an eigenvalue bifurcating from $\pm\lambda_{1-}$ that will move towards the origin and later becomes a pair of real eigenvalues. These solitons are also unstable in the continuum limit.

\begin{figure}[htbp!]
\centering
{\includegraphics[width=7cm]{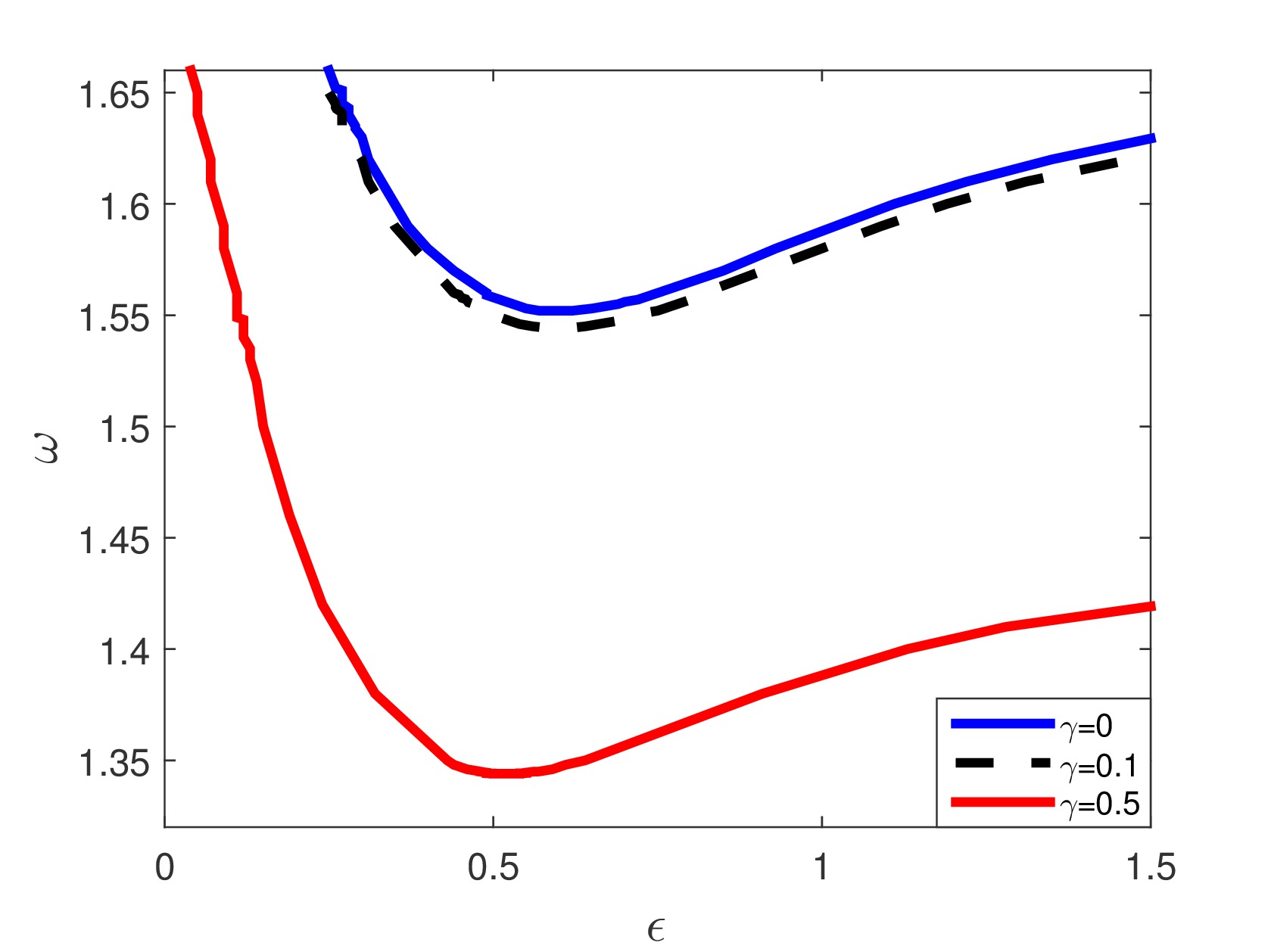}}
{\includegraphics[width=7cm]{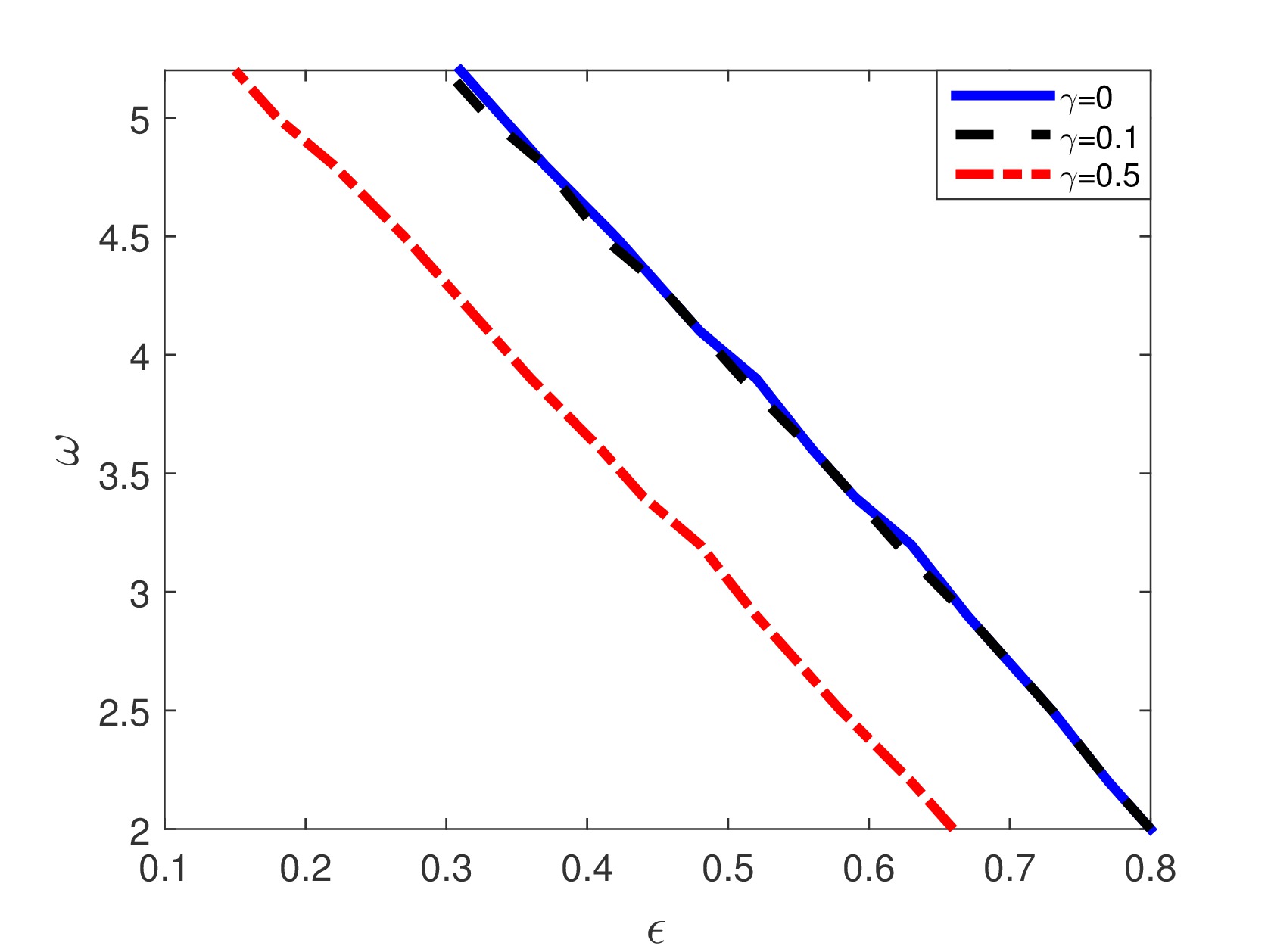}}
\caption{The stability region of the onsite soliton type I (left) and II (right) in the $(\epsilon,\omega)$-plane for several values of $\gamma$. The solutions are unstable above the curves.}
\label{si2}
\end{figure}

Unlike intersite discrete solitons that are always unstable, onsite discrete solitons may be stable. In Fig.\ \ref{si2}, we present the (in)stability region of the two types of discrete solitons in the $(\epsilon,\omega)$-plane for three values of the gain-loss parameter $\gamma$. Discrete solitons are unstable above the curves. Indeed as we mentioned before, for soliton I there is a critical $\omega$ that depends on $\gamma$ below which the soliton is stable in the continuum limit, while soliton II is always unstable in that limit. Another difference between the two figures is that the stability curves in the left panel generally corresponds to an eigenvalue crossing the origin that becomes real-valued, while the curves in the other panel are due to the appearance of a quartet of complex eigenvalues. In general, we obtain that the gain-loss term can be parasitic as it reduces the stability region of the discrete solitons.

\begin{figure}[htbp!]
\centering
{\includegraphics[width=5cm]{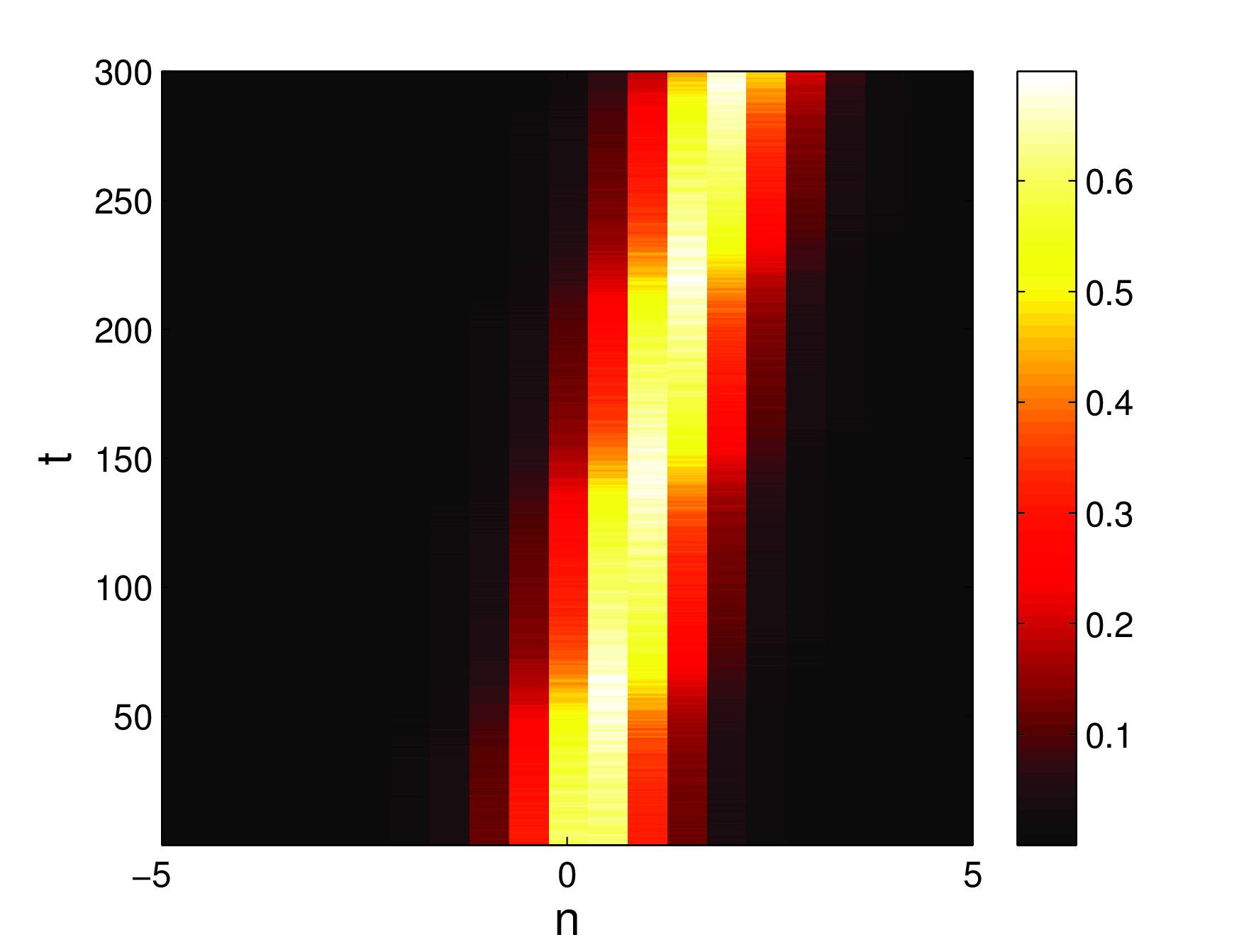}}
{\includegraphics[width=5cm]{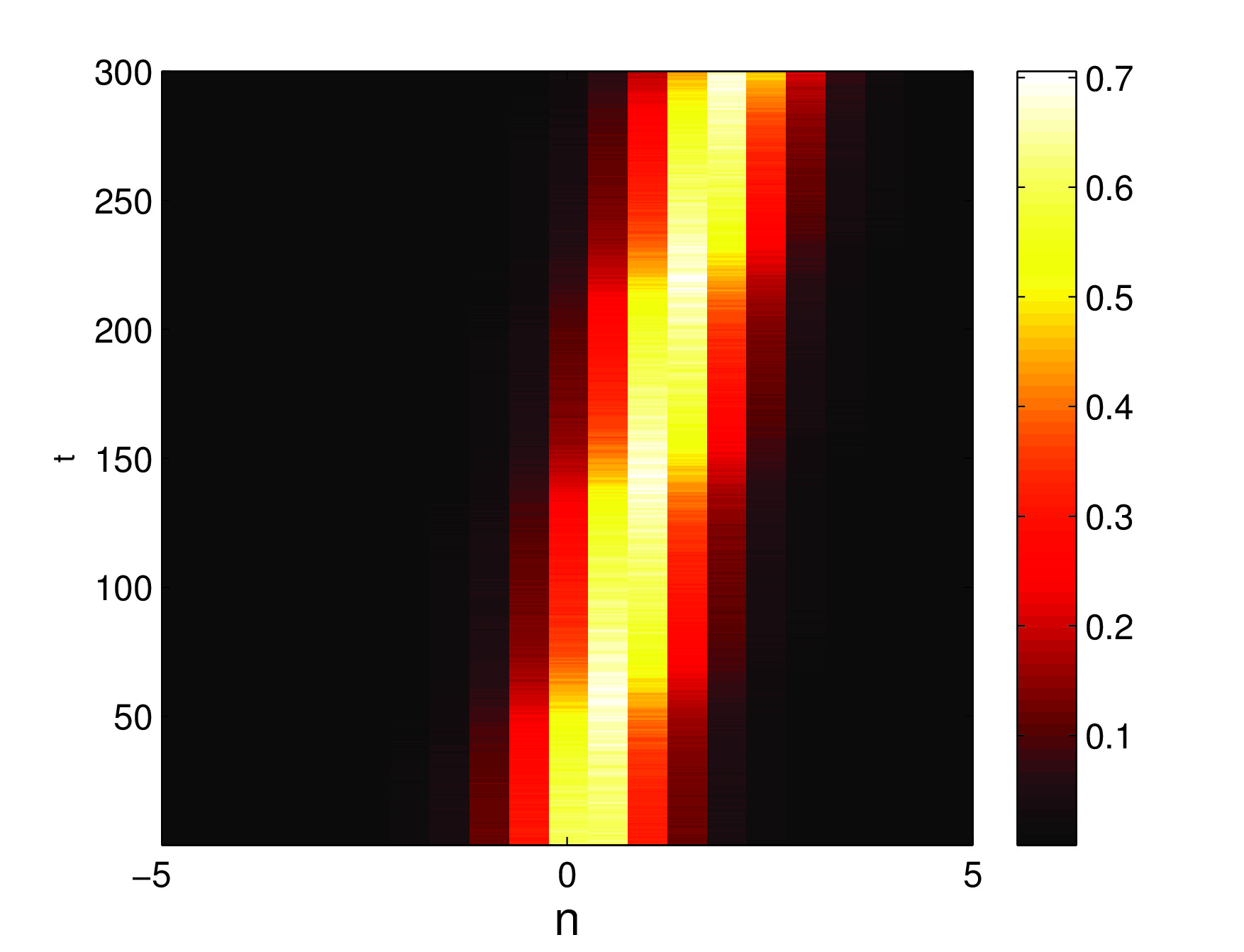}}\\
{\includegraphics[width=5cm]{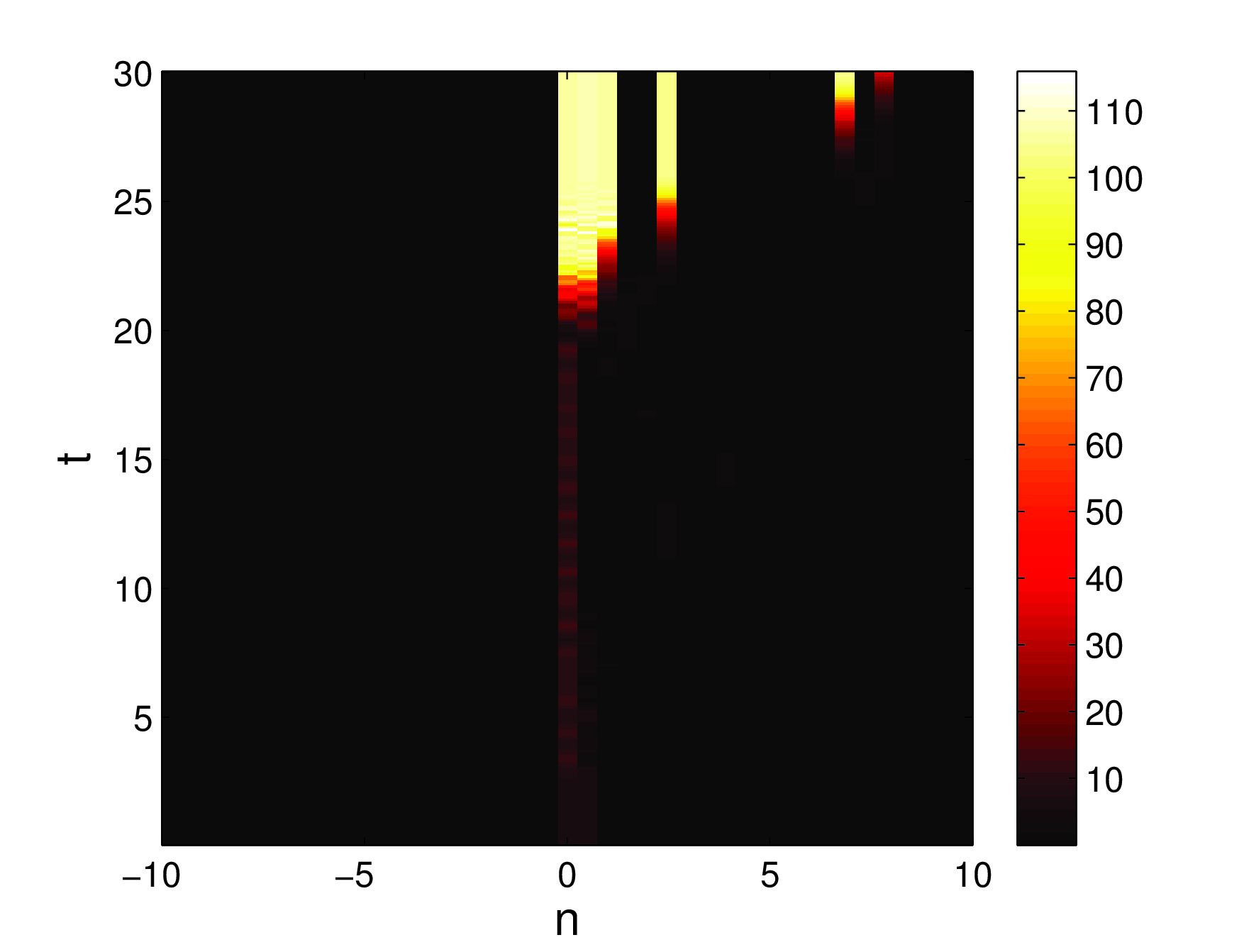}}
{\includegraphics[width=5cm]{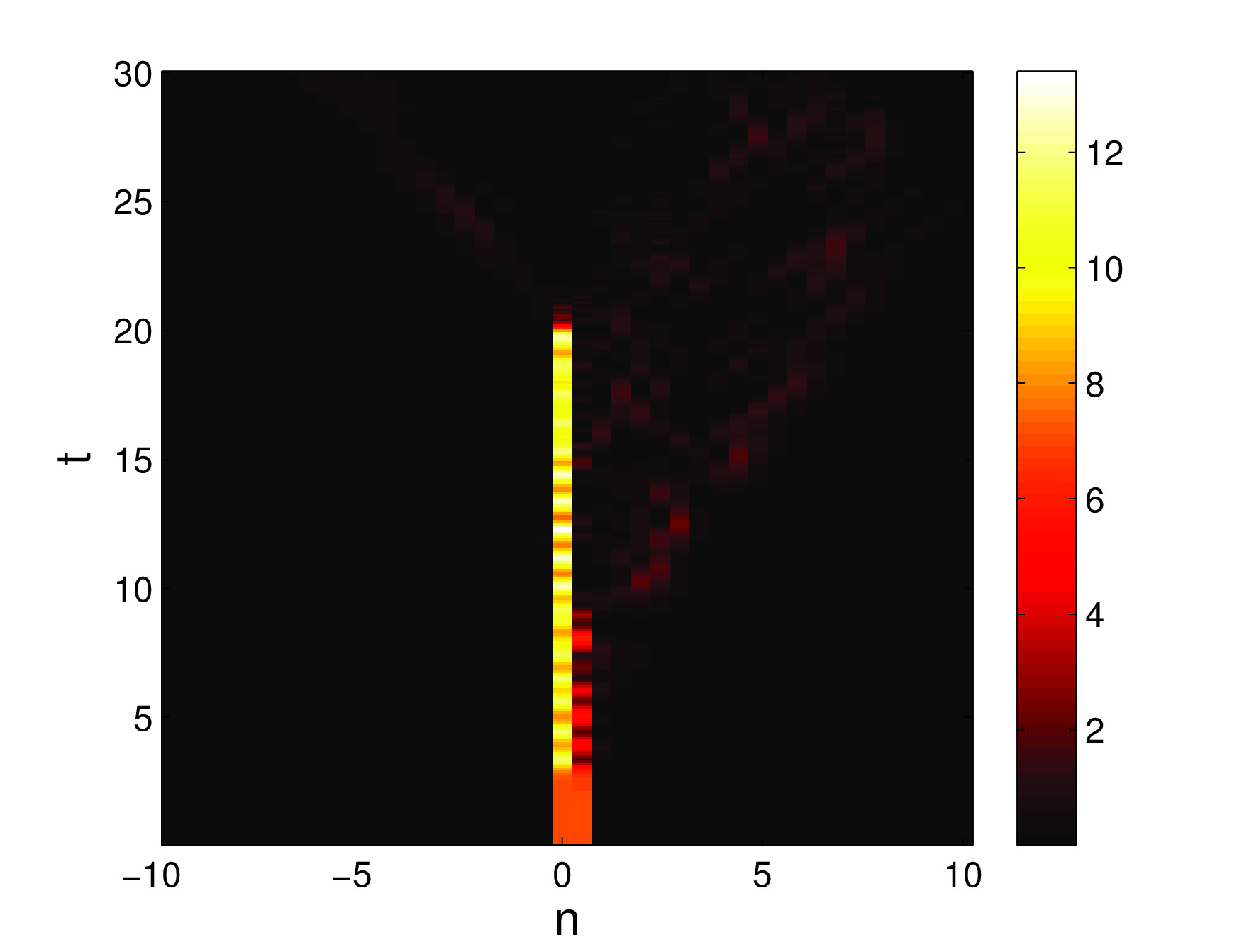}}\\
{\includegraphics[width=5cm]{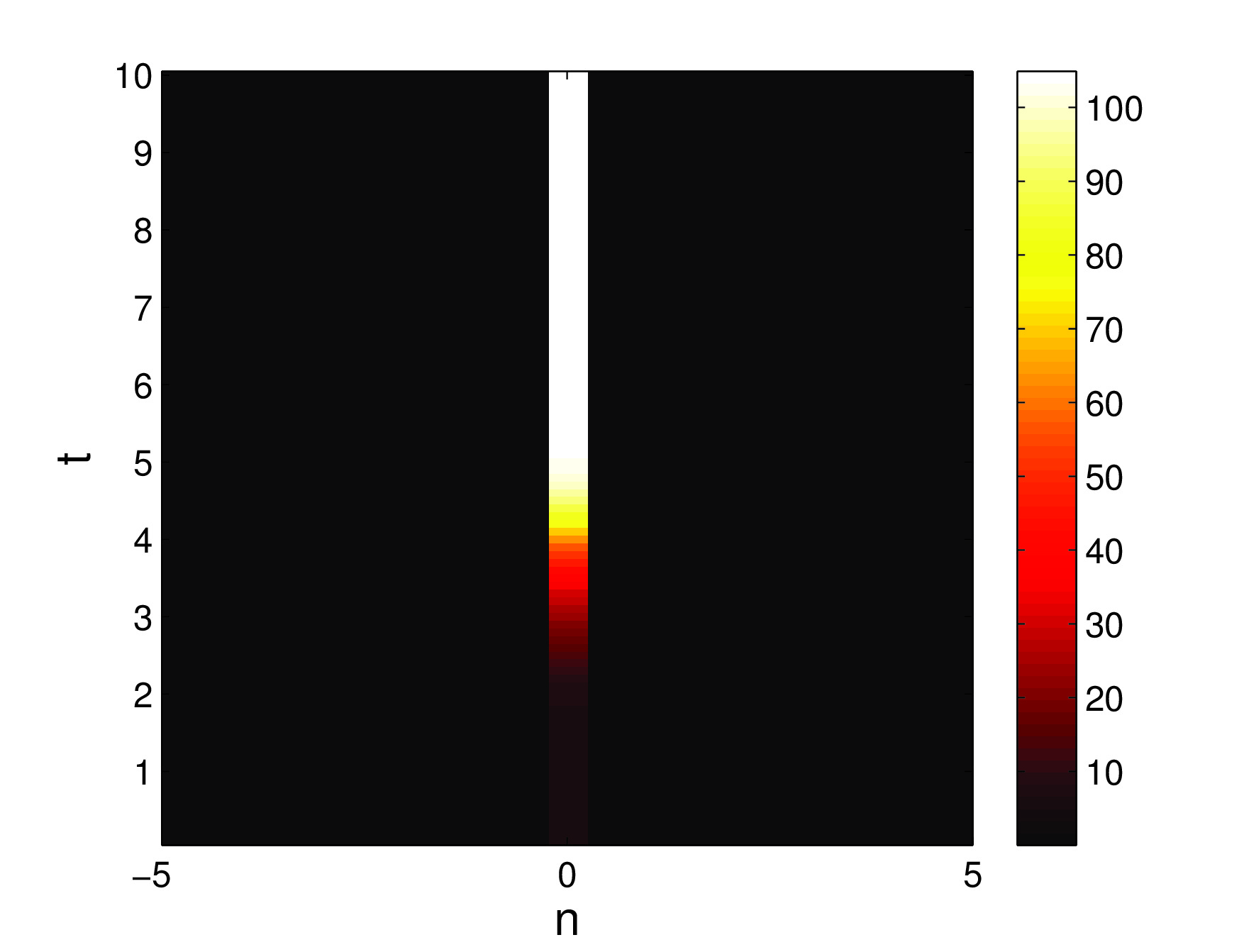}}
{\includegraphics[width=5cm]{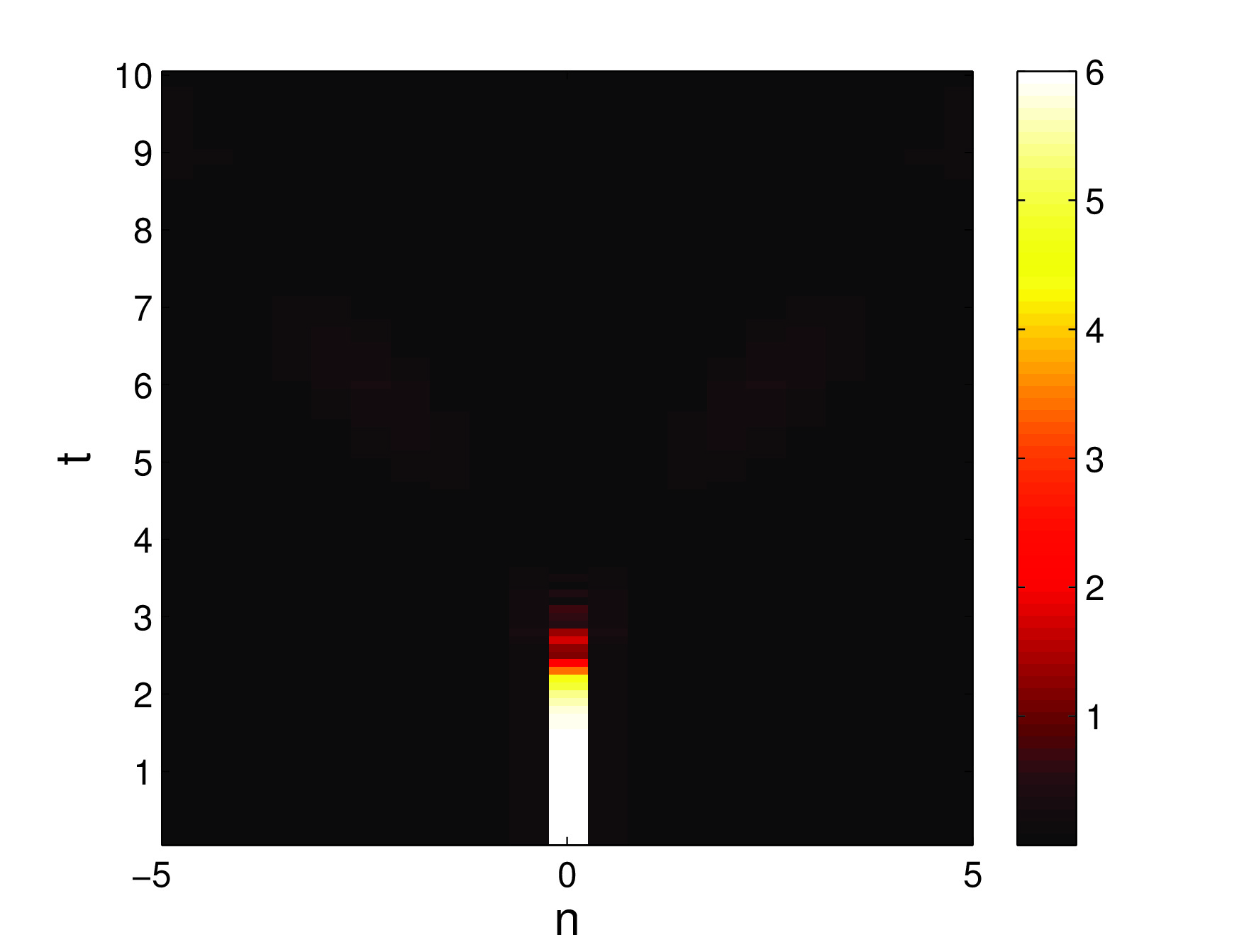}}\\
{\includegraphics[width=5cm]{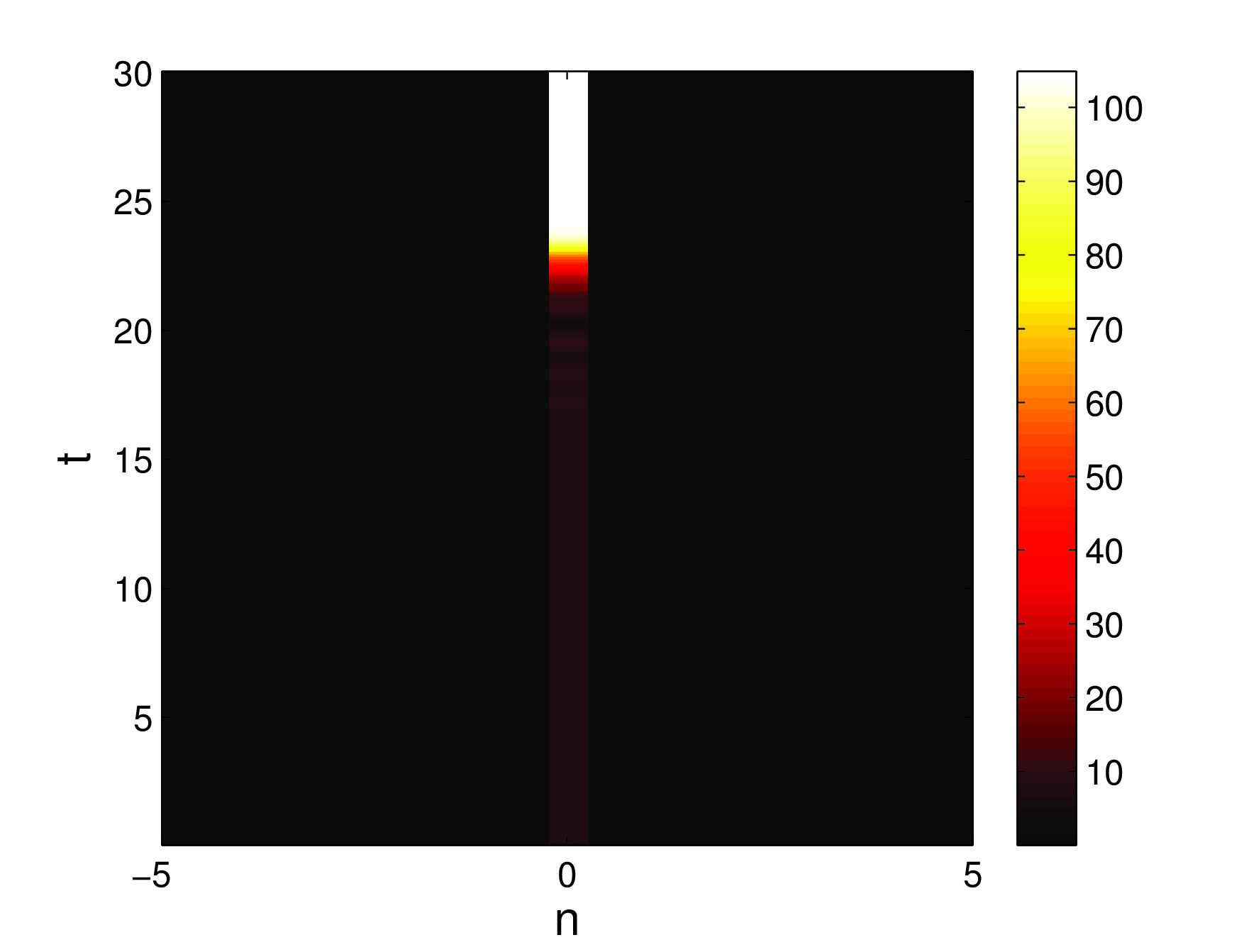}}
{\includegraphics[width=5cm]{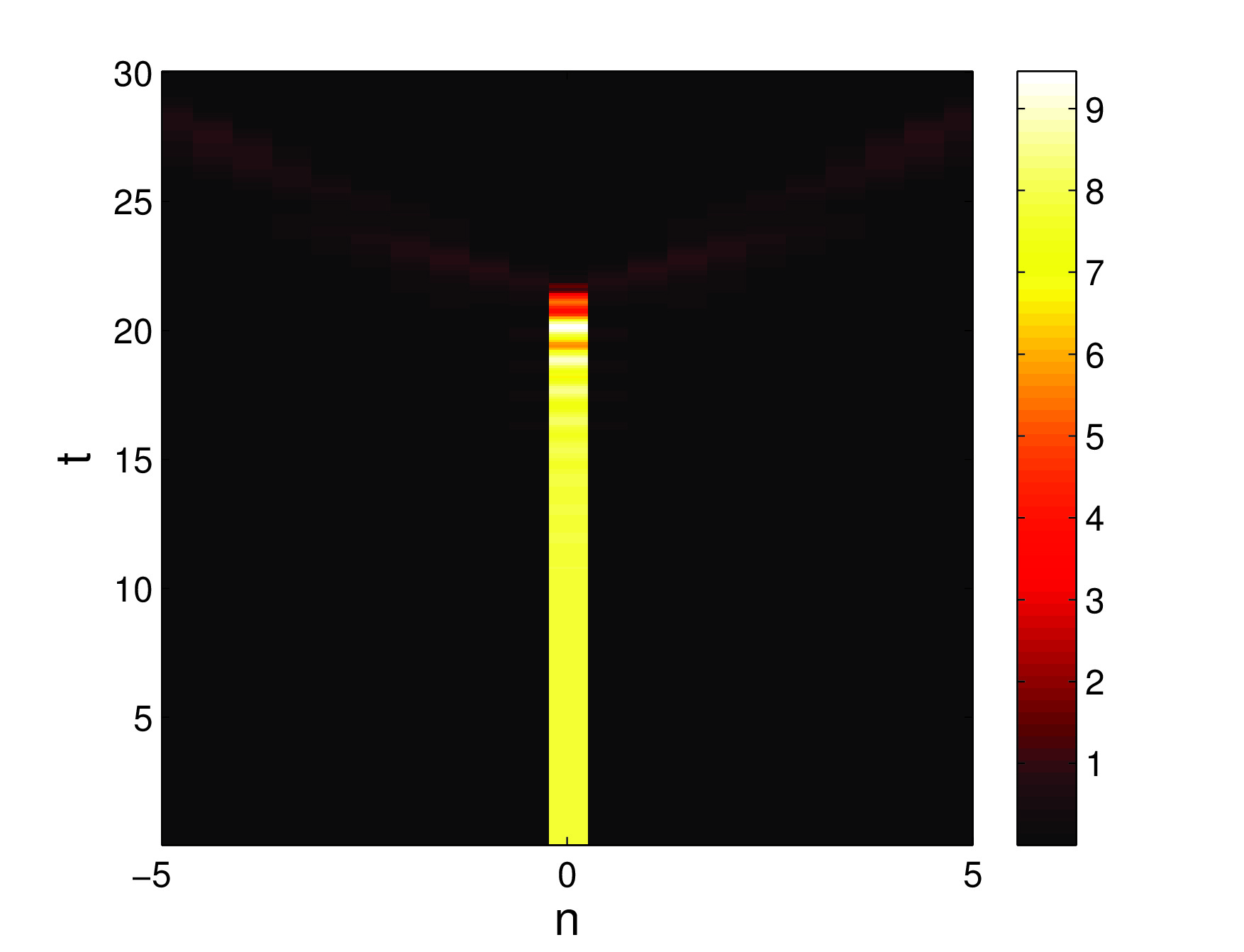}}
\caption{The typical dynamics of the instability of the discrete solitons in the previous figures. Here, $\gamma=0.5$ and $\epsilon=1$. Depicted in the left and right panels are $|u_n|^2$ and $|v_n|^2$, respectively. From top to bottom panels, shown are the dynamics of intersite soliton I with $\omega=1.2$, intersite soliton II with $\omega=5.2$, and onsite soliton I and II both with $\omega=5.2$.}
\label{dyn}
\end{figure}

Finally, we present in Fig.\ \ref{dyn} the time dynamics of some of the unstable solutions shown in the previous figures. What we obtain is that typically there are two kinds of dynamics, i.e.\ in the form of travelling discrete solitons or solution blow-ups. The first type was the typical dynamics of the intersite soliton I. The second dynamics is typical for the other types of unstable discrete solitons.

\section{Conclusion}

We have presented a systematic method to determine the stability of discrete solitons in a \pts-symmetric coupler by computing the eigenvalues of the corresponding linear eigenvalue problem using asymptotic expansions. We have compared the analytical results that we obtained with numerical computations, where good agreement is obtained. From the numerics, we have also established the mechanism of instability as well as the stability region of the discrete solitons. The application of the method in higher dimensional \pts-symmetric couplers (see, e.g., \cite{chen14}) is a natural extension of the problem that is addressed for future work. Additionally we also address the computation of eigenvalues of discrete solitons in the neighbourhood of broken \pts-symmetry as future investigations. 

\section*{Acknowledgement}

AAB and HS are grateful to the University of Nottingham for the 2013 Visiting Fellowship Scheme and British Council for the 2015 Indonesia Second City Partnership Travel Grant.\\

\noindent The authors declare that there is no conflict of interest regarding the publication of this paper.

\appendix

\section{Analytical calculation}

As mentioned in Section \ref{sds}, to solve the eigenvalue problem (\ref{8}) analytically we expand the eigenvalue and eigenvector asymptotically as
\begin{equation}
\Box = \Box^{(0)}+\sqrt\epsilon \Box^{(1)}+\epsilon\Box^{(2)}+\dots,
\label{box1}
\end{equation}
with $\Box=\lambda, K_n, L_n, P_n, Q_n$.

Performing the expansion in $\epsilon$, 
at $\mathcal{O}(\epsilon^0)$ we obtain the following set of equations
\begin{eqnarray}
\fl\lambda^{(0)}
\underline{v}_n^{(0)} &=&
\underbrace{\left[{\begin{array}{*{20}c}
\gamma & \omega-{A_{n}^{(0)}}^2 & 0 & -1 \\ 3(A_{n}^{(0)})^2-\omega &\gamma& 1 &0 \\0  & -1& k_1 & k_2\\1&0&k_3&k_4
\end{array}}\right]}_{M_0}
\underline{v}_n^{(0)},\label{e0}
\end{eqnarray}
where
\begin{eqnarray*}
& \underline{v}_n^{(j)}=\left[{\begin{array}{c}
K_{n}^{(j)} \\ L_{n}^{(j)} \\ P_{n}^{(j)}\\ Q_{n}^{(j)} \end{array}}\right], &\\ &k_1=-(2\Re (B_{n}^{(0)})\Im (B_{n}^{(0)})+\gamma),\, &k_2=\omega-\Re (B_{n}^{(0)})^{2}-3\Im (B_{n}^{(0)})^{2},\\ &k_3=3\Re (B_{n}^{(0)})^{2}+\Im (B_{n}^{(0)})^{2}-\omega, &k_4=2\Re (B_{n}^{(0)})\Im (B_{n}^{(0)})-\gamma.
\end{eqnarray*}

At $\mathcal{O}(\epsilon^{1/2})$ and $\mathcal{O}(\epsilon^{1})$, we obtain
\begin{eqnarray}
\lambda^{(0)}\underline{v}^{(1)}_n &&=
M_0\underline{v}_n^{(2)}-\lambda^{(1)} \underline{v}^{(0)}_n, \label{e1}\\
\lambda^{(0)}M_0 &&= M_0\underline{v}_n^{(2)}
-\lambda^{(1)}\underline{v}_n^{(1)}
-\lambda^{(2)}\underline{v}_n^{(0)}\nonumber\\
&&+\underbrace{\left[{\begin{array}{*{20}c} 0 & 2\left(1-A_{n}^{(0)}A_{n}^{(1)}\right) & 0 & 0 \\ 6A_{n}^{(0)}A_{n}^{(1)}-2 &0& 0 &0 \\0  & 0& k_5 & 2+k_6\\0&0&k_7-2&k_8 \end{array}}\right]}_{M_1}
\underline{v}_n^{(0)}\nonumber\\
&&+\left[{\begin{array}{*{20}c} 0 & -1& 0 & 0 \\ 1 &0& 0 &0 \\0  & 0& 0 & -1\\0&0&1&0 \end{array}}\right]
\left(\underline{v}_{n+1}^{(1)}+\underline{v}_{n-1}^{(1)}\right),\label{e2}
\end{eqnarray}
where
\begin{eqnarray*}
&k_5=-2(\Re (B_{n}^{(0)})\Im (B_{n}^{(1)})+\Re (B_{n}^{(1)})\Im (B_{n}^{(0)})),\\&k_6=-2(\Re (B_{n}^{(0)})\Re(B_{n}^{(1)})+3\Im (B_{n}^{(0)})\Im(B_{n}^{(1)})),\\
&k_7=2(3\Re (B_{n}^{(0)})\Re(B_{n}^{(1)})+\Im (B_{n}^{(0)})\Im(B_{n}^{(1)})),\\&k_8=2(\Re (B_{n}^{(0)})\Im (B_{n}^{(1)})+\Re (B_{n}^{(1)})\Im(B_{n}^{(0)})).
\end{eqnarray*}

The steps of finding the coefficients $\lambda^{(j)}$ of the asymptotic expansions, $j=0,1,2,\dots$, are as follows.
\begin{enumerate}
\item Solve the eigenvalue problem (\ref{e0}), which is a $4\times4$ system of equations, for $\lambda^{(0)}$ and $\underline{v}_n^{(0)}$.
\item Determine $\lambda^{(1)}$ by taking the vector inner product of both sides of (\ref{e1}) with the null-space of the Hermitian transpose of the block matrix that consists of $\left( M_0-\lambda^{(0)}I_4\right)$ along the diagonal, where $I_4$ is the $4\times4$ identity matrix.
\item Solve (\ref{e1}) for $\underline{v}_n^{(1)}$.
\item Determine $\lambda^{(2)}$ by taking the vector inner product of both sides of (\ref{e2}) with the null-space of the Hermitian transpose of the block matrix that consists of $\left(M_0-\lambda^{(0)}I_4\right)$.
\end{enumerate}
The procedure repeats if one would like to calculate the higher order terms.

The leading order eigenvalue $\lambda^{(0)}$ of (\ref{e0}) has been solved in \cite{li11}. However, the expression of the corresponding eigenvector $\underline{v}_n^{(0)}$ was very lengthy, that makes it almost impractical to be used to determine the higher order corrections of $\lambda^{(j)}$. Therefore, in every equation at order $\mathcal{O}(\epsilon^\ell)$ obtained from (\ref{8}), we also expand the variables in $\gamma$, i.e.\
\[
\fl M_j=M_{j,0}+\gamma M_{j,1}+\gamma^2 M_{j,2}=\dots,\,\Box^{(j)}=\Box^{(j,0)}+\gamma\Box^{(j,1)}+\gamma^2\Box^{(j,2)}+\dots,
\]
where again $\Box=\lambda, \underline{v}_n$, and obtain equations at order $\mathcal{O}(\gamma^{\hat{\ell}})$. The steps to determine $\lambda^{(j,k)}$ and $\underline{v}_n^{(j,k)}$ are the same as mentioned above.

\end{document}